# Wireless Optogenetic Nanonetworks: Device Model and Charging Protocols

Stefanus A. Wirdatmadja, Michael T. Barros, Yevgeni Koucheryavy, Josep Miquel Jornet, Sasitharan Balasubramaniam

***Abstract*—In recent years, numerous research efforts have been dedicated towards developing efficient implantable devices for brain stimulation. However, there are limitations and challenges with the current technologies. Firstly, the stimulation of neurons currently is only possible through implantable electrodes which target a population of neurons. This results in challenges in the event that stimulation at the single neuron level is required. Secondly, a major hurdle still lies in developing miniature devices that can last for a lifetime in the patient's brain. In parallel to this, the field of optogenetics has emerged where the aim is to stimulate neurons using light, usually by means of optical fibers inserted through the skull. Obviously, this introduces many challenges in terms of flexibility to suit the patient's lifestyle and biocompatibility. We have recently proposed the concept of *wireless optogenetic nanonetworking devices (WiOptND)* that could address the problem of long term deployment, and at the same time target single neuron stimulation [1]. The WiOptND is equipped with a miniature LED that is able to stimulate the genetically engineered neuron, and at the same time harvest energy from ultrasonic vibrations. This paper investigates the behaviour of light propagation in the brain tissue, and based on the power required to emit sufficient intensity for stimulation, a circuitry for the energy harvesting process is designed. In addition, a number of charging protocols are also proposed, in order to minimize the quantity of energy required for charging, while ensuring minimum number of neural spike misfirings. These protocols include the simple *Charge and Fire*, which requires the full knowledge of the raster plots of neuron firing patterns, and the *Predictive Sliding Detection Window*, and its variant *Markov-Chain based Time-Delay Patterns*, which minimizes the need for full knowledge of neural spiking patterns. Moreover, extensive simulation results are conducted to compare the performance of the different charging protocols. The drop of stimulation ratio for ~25% and more stable trend in its efficiency ratio (standard deviation of ~0.5%) are exhibited on *Markov-Chain based Time-Delay Patterns* protocol compared to *Change and Fire*. The results show the feasibility of utilizing WiOptND for long-term implants in the brain, and a new direction towards precise stimulation of neurons in the cortical column of the brain cortex.***

## I. Introduction

Each year, the prevalence of neurodegenerative diseases, such as Alzheimer's disease, amyotrophic lateral sclerosis and Parkinson's disease, is increasing. According to the 2016 *World Health Organization* (*WHO*) data statistics, more than five million Americans are living with Alzheimer's and it is predicted that the number will increase to around 16 million by 2050 [2]. Parkison's affects 500,000 people in the US and it will double by 2030 [3]. It has an estimated cost of 20 billion dollars in the US [4] and 13 billion euros in Europe [5]. This situation demands scientists and researchers to not only develop prevention programs, but also solutions that might assist the patients to live a normal lifestyle. In certain cases, patients with Parkinson's disease may receive benefits from brain stimulation by placing *Implantable Pulse Generator (IPG)* [6] to the targeted areas of the brain to treat essential tremor and dystonia symptoms. In the field of neuroscience, *optogenetics* is gaining popularity as an alternative method for neural stimulation. In optogenetics, the neurons are engineered so that they are sensitive to light at specific wavelengths in order to have either excitatory or inhibitory effects [7]. Optogenetics provides an advantage over the use of electrodes due to its higher precision, less stress to the cells, and lower noises. These noises may disturb the neural activity recording process since the recorded signal does not merely come from the target neurons, but also from the stimulation electrodes [8]. However, further improvements are still required to the current solutions. They include the degree of intrusion through the skull and alternative power supply compared to the use of batteries. These shortcomings limit the degree of practicality for the patients in their daily life.

On neural networks, neurons communicate with each other through the process of action potentials and synapses. The neuron cell body (soma) connects to other cells using the dendrites and axons, which receives and transmits signals with neighbouring cells. The problem arises when this physical transportation network is impaired due to various problem such as aging, disease, or the death of cells, among others. Even a single neuron failure may result in communication impairments along the cortical circuit. This communication impairment due to single cell level failure will result in the discontinued transmission of action potential among cortical layers as depicted in Fig. 1. In this case, the implementation of a single neuron level stimulation implant can restore the neural circuit communication between the layers.

We have recently proposed the concepts of wireless optogenetics integrated using nanoscale components [1], which we term *wireless optogenetic nanonetworking devices* (*WiOptND*). Considering the size of the soma which varies between 4-100$\mu m$ in diameter, the WiOptND is required to be approximately several hundreds microns in size in order to deliver the required light intensity needed for the stimulation. Since our aim is to embed the device into the brain, this means that this miniaturization will require a light source

S. A. Wirdatmadja, S. Balasubramaniam and Y. Koucheryavy are with Department of Electrical Engineering and Communications, Tampere University of Technology, Finland. Email: (stefanus.wirdatmadja, sasi.bala, yk)@tut.fi

S. Balasubramaniam and M. T. Barros is with the Telecommunication Software & Systems Group (TSSG) at Waterford Institute of Technology (WIT), Ireland. e-mail: mbarros@tssg.org

J. Jornet is with the Department of Electrical Engineering, University at Buffalo, The State University of New York, US. Email: jmjornet@buffalo.edu

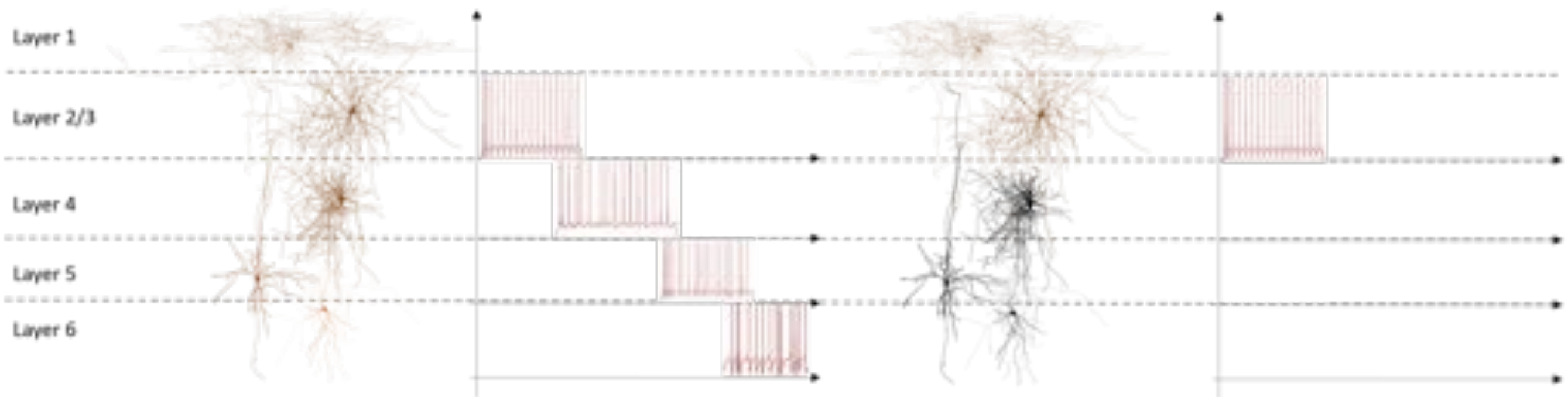

Fig. 1: Illustration of healthy and disconnected cortical neural networks. Failing of action potential relays will result in disconnected communication in the cortical column of the cerebral cortex.

that can emit sufficient light intensity, and is powered using energy harvesting [1] [9]. In this paper, we investigate the light propagation behaviour through the tissue to determine the effective distance required between the light source and the neurons, and the energy harvesting technique based on ultrasound vibrations of piezoelectric nanowires.

In addition, since we anticipate a network of WiOptNDs, a suitable charging protocol is required to maximize the energy efficiency and ensure that all devices are charged to minimize misfiring of the neurons. In particular, since these devices will be embedded into the cerebral cortex, the charging protocol should consider the characteristics of the neural circuit, and in particular, knowledge of the neural spike sequence. We propose three protocols, including the simple *Charge and Fire* protocol based on utilizing the full knowledge of the raster plot firing sequence to individually charge each device in the network, the *Predictive Sliding Detection Window* protocol which employs the combination of neural spike prediction and a more complex circuit for parallel charging in order to minimize usage of ultrasound frequencies, and its variant the *Markov-Chain based Time-Delay Patterns*, which utilizes knowledge of the neuron population and their connection probability between the cortical layers to predict the firing sequence to also minimize the required ultrasound frequencies.

The structure of the paper is as follows: background information on optogenetics and its biological features are provided in Sec. II. Going more into the WiOptND, the overall model and the components are discussed in Sec. III. Light propagation properties through the brain tissue is discussed in Sec. IV, while in Sec. V the power management of WiOptND is presented, particularly focusing on the energy charging and releasing performance to power the light source. The charging protocols of the WiOptND nanonetwork is discussed in Sec. VI. Finally, the paper is concluded in Sec. VII.

## II. Optogenetics

Naturally, the communication between neurons is done both electrically and chemically. In most cases, the electric signal is used in transferring the information in one single neuron, while chemical is used in inter-neuronal communications [10] [11]. The electrical stimulation, action potential, propagates from dendrites to the axons and stimulates the release of neurotransmitter for inter-neuronal synapse communications. The most common method in controlling the neural communications is using electrical and light stimulation (optogenetics). Comparing both methods, optogenetics gives higher precision in targeting specific neuron, minimizes the cell stress as compared to electrical stimulation, and creates less interference with the surrounding cells.

Optogenetics is a method of artificially manipulating neural communication using light at a specific wavelength. According to its characteristics, the optogenetic construct can have either excitatory or inhibitory effects. *Excitatory postsynaptic potential (EPSP)* refers to the case when the cell membrane depolarizes as a result of the opening of sodium and calcium ion membrane channels, causing action potential to be generated. On the other hand, *inhibitory postsynaptic potential (IPSP)* is when the cell membrane hyperpolarizes caused by the opening of chloride or potassium ion membrane channels which results in blockage of action potential generation.

In optogenetics, *channelrhodopsin-2* (*ChR2*) exhibits excitatory characteristics. This construct is obtained by genetically engineering neurons with the *opsins* from green algae *Chlamydomonas reinhardtii* (Step 1 in Fig. 2) [12]. The blue light illumination triggers the action potential generation (Step 2 and 3 in Fig. 2). For inhibitory effect, the hyperpolarization can be done in two ways, using either chloride or proton pumps. Chloride pump is realized by utilizing the *halorhodopsin* (*NpHR*) from archaeon *Natronomonas pharaonis* [13]. The improved version of NpHR is called eNpHR3.0 which can be activated by green, yellow, or red light (Step 2 and 3 in Fig. 2). During its activation, chloride ion channel gates open, bringing chloride ions into the cells. Proton pump is the alternative of chloride pump to perform the inhibitory effect. To create proton pumps, there are four types of optogenetics that can be used. They are archaerhodopsin-3 (Arch) from *Halorubrum sodomense*, Mac from the fungus *Leptosphaeria maculans*, archaerhodopsin (ArchT) from *Halorubrum strain TP009*, and eBR (an enhanced version of bacteriorhodopsin from *Halobacterium salinarum*) [14]. In a nutshell, Fig. 2 concludes how the implementation of ChR2 and NpHR affects the generation of action potentials upon the illumination of blue light (480 $nm$) and yellow light (570 $nm$).

Since the focus of this paper is to excite the neurons, the optogenetic construct that is used is the ChR2, which is activated by blue light whose wavelength is approximately 480 nm.

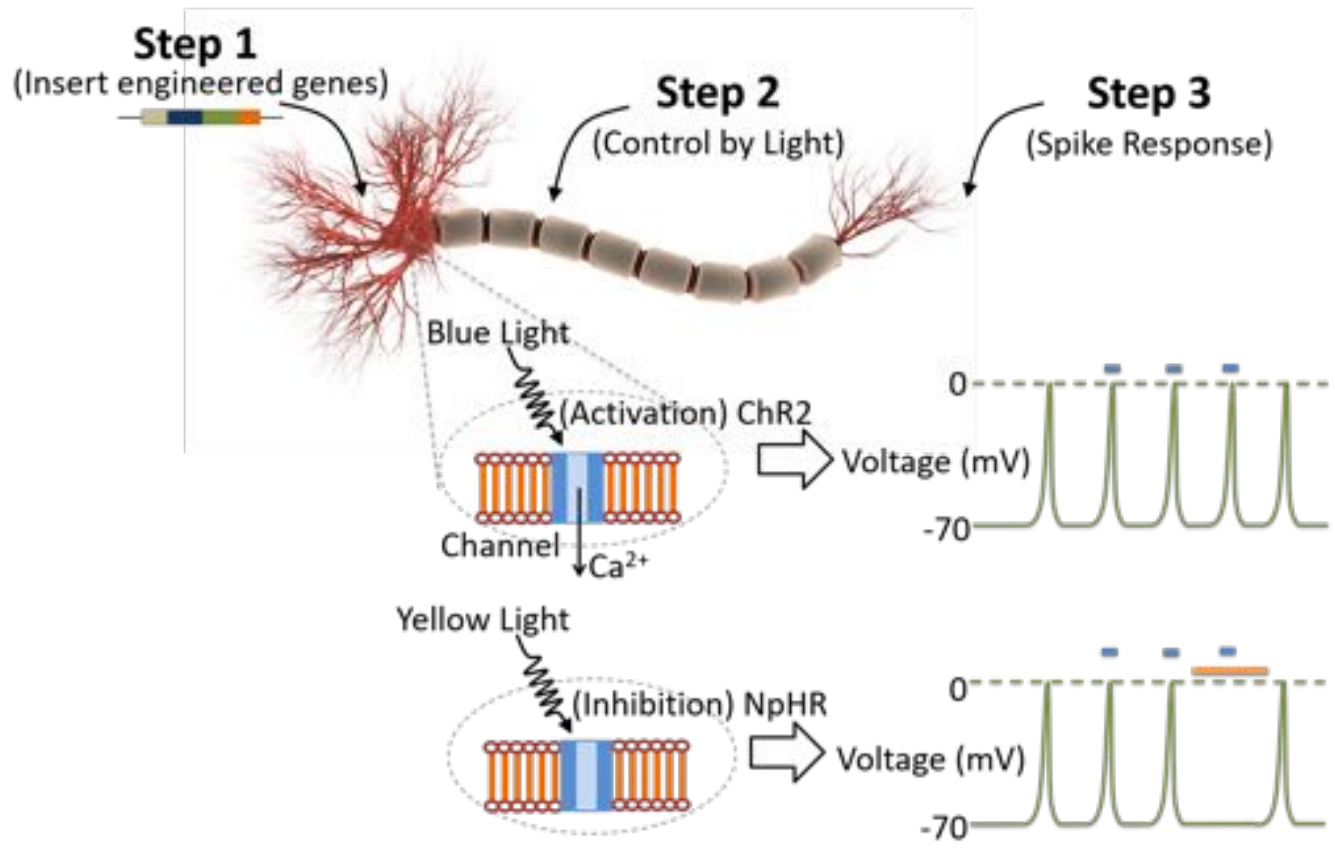

Fig. 2: Steps towards developing an optogenetic neuron and its stimulation process. The figure also illustrates the depolarization process where ion pumps are activated.

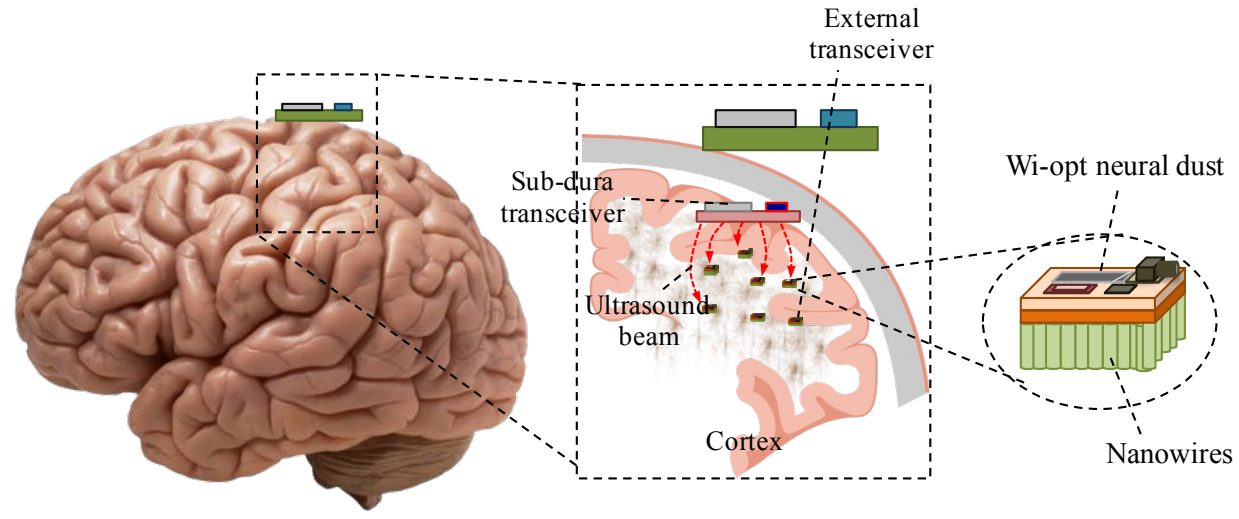

Fig. 3: Illustration of the overall architecture of the Wireless Optogenetic Nanonetwork. The WiOptND are scattered in the various layers of the cortex, and is charged by the ultrasound signals emitted from the *sub-dura transceiver*, which in turn is communicated from the *external transceiver*.

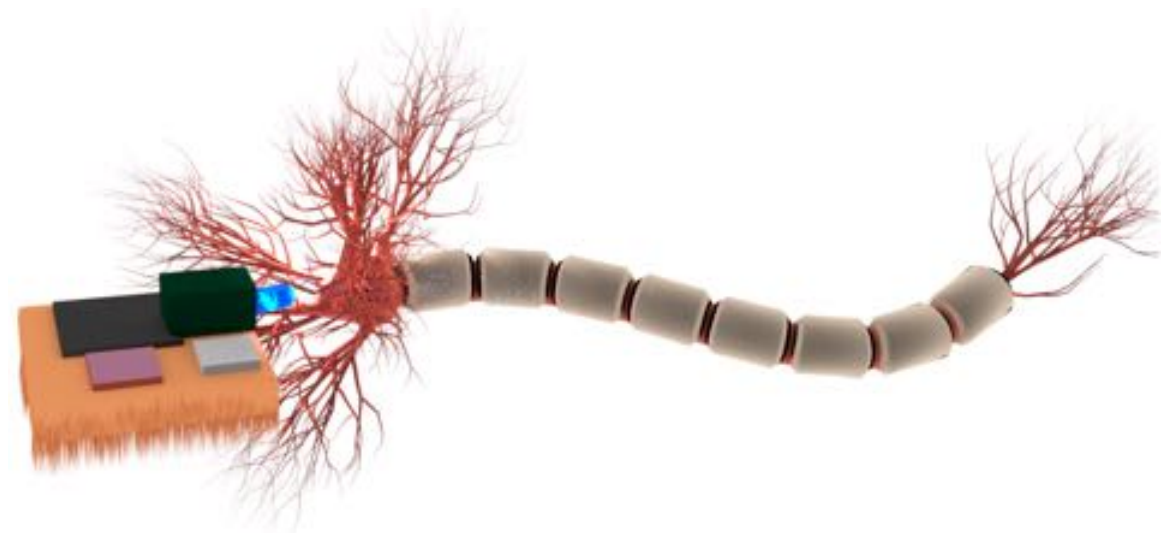

Fig. 4: Illustration of a WiOptND that interfaces to an engineered neuron that is sensitive to light at a specific wavelength.

## III. System Model

This section will first describe the devices for each of the layers of the architecture, and this will be followed by the device design, including the different components of the circuitry.

### A. Wireless Optogenetic Nanonetworking Architecture

The entire network of the Wireless Optogenetic Nanonetworking architecture is composed of three layers, which is adopted from [15]. The lowest layer is the cerebral cortex where the WiOptNDs are distributed, each are interfaced to a neuron that requires stimulation as illustrated in Fig. 4. Cerebral cortex is the gray matter of the brain and is responsible for sensory, motor, and associated functions. Horizontally, the cerebral cortex is categorized based on their functional areas, while vertically, it comprises of six layers containing different type of neurons, which includes: *pyramidal cells*, *spiny stellate cells*, *basket cells*, *chandelier cells*, and *smooth stellate cells* [16], each of which can have a WiOptND interfaced to the cell. The next layer up is the *sub-dura transceiver*, which is located on the dura and below the skull, and communicates with the WiOptND. The role of the sub-dura transceiver is to emit ultrasound waves which are used to charge the WiOptND. The sub-dura transceiver contains the algorithm the determines both the charging and stimulation sequence of the WiOptND, and this in turn emits the sequence of ultrasound signals. Above the sub-dura transceiver is the *external transceiver*, which communicates with the sub-dura transceiver (please note that this paper does not focus on the interactions between the *external transceiver* and the *sub-dura transceiver*).

### B. Wireless Optogenetic Nanonetworking Devices

The circuit diagram of the WiOptND is illustrated in Fig. 5. Acting as the energy harvester, the *piezoelectric* nanowires vibrate in response to the radiated ultrasound wave emitted from the sub-dura transceiver. As the nanowires oscillate, the AC voltage is generated. In this stage, the mechanical energy from the ultrasound pressure is converted into electrical energy in accordance to the piezoelectric material. However, the ultrasound intensity must abide with the Food and Drug Administration (FDA) safety regulation stating that the maximum permissible level is 720 $mW/cm^2$. The generated AC voltage is converted to DC by using a *rectifier*. The converted electrical energy is then stored in the storage capacitor which is charged to power the light source component, and in our case is a *Light Emitting Diode* (*LED*). Since our aim is to be able to signal each individual WiOptND in order to charge and trigger the stimulation process, a unique addressing scheme is required for each device. One approach is through the utilization of a *Voice Operated Switch* (*VOX*) that responds to different ultrasound frequencies. By integrating the frequency filter switch, adopted from the VOX, enables the discharging selection of one particular device, and this is achieved by integrating a piezo element that is sensitive to specific resonant frequency.

## IV. Light Propagation in Brain Tissue

As the light emitted from the LED traverses via the brain tissue, the irradiance or the intensity decreases. Absorption due to the tissue *chromophores* increases as the light is scattered along its propagation path. The main chromophores in biological tissue include water, lipids, melatonin, oxygenated and deoxygenated haemoglobin. Eighty percent of an average human brain contains water, however, its absorption coefficient is negligible especially for visible light. The same phenomenon occurs for lipids as well, since the lipid content is approximately 5% of the brain [17]. The large percentage of the brain comprises blood which delivers oxygen from the lungs to the brain and vice versa. Haemoglobin of red blood cells contains the highest light absorbing chromophores.

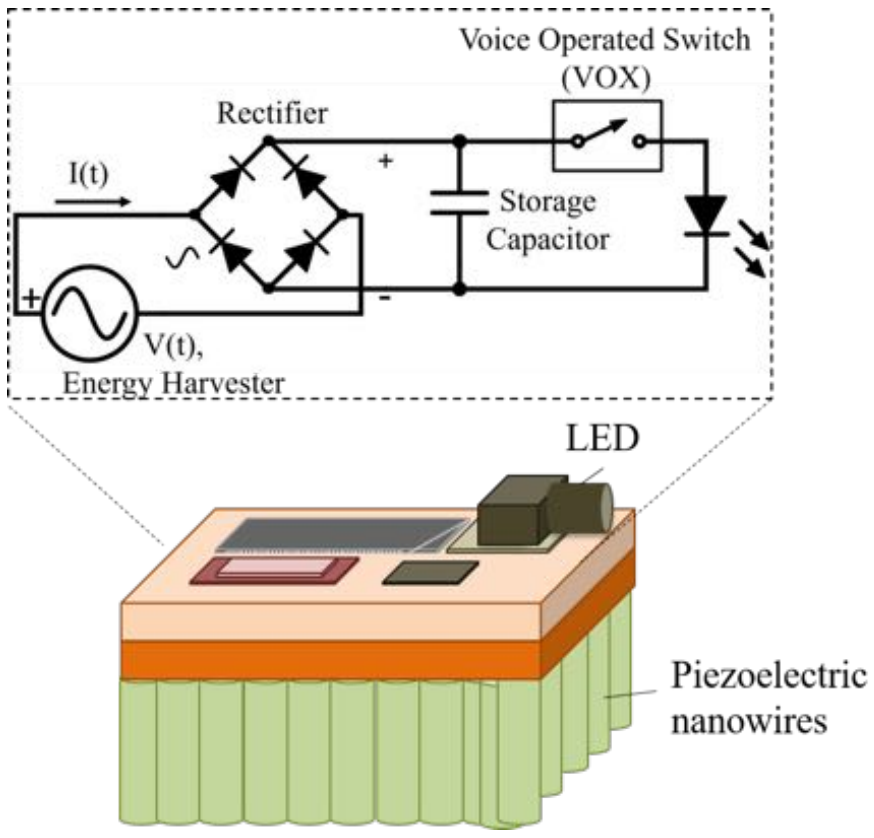

Fig. 5: Device architecture of the WiOptND, including the internal circuit diagram.

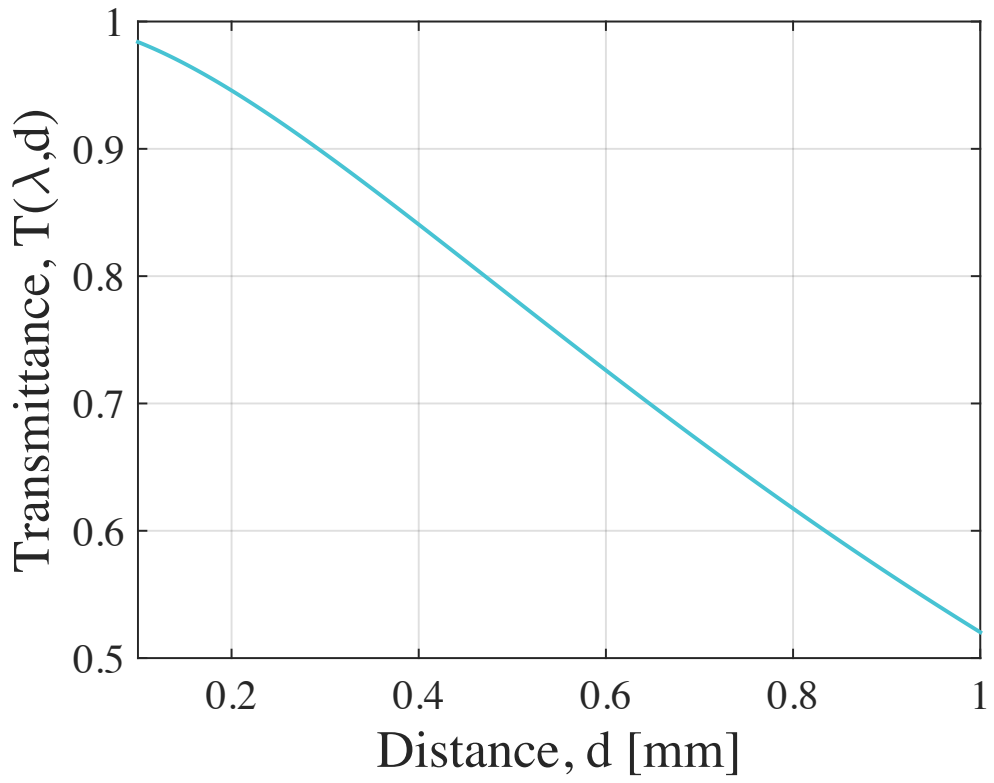

Fig. 6: Intensity ratio as a function of distance from the light source. The curve is greatly affected by the light absorption and scattering phenomenon in the brain tissue.

For the light intensity requirement analysis for ChR2 activation, the *Modified Beer-Lambert Law* can be used for photon transport modelling. According to the differential form of this model, the absorption in the tissue is proportional to the major chromophores concentration. Furthermore, this model assumes constant scattering losses and homogeneous semi infinite medium (tissue) reflected on the absorption factor. The *Differential Pathlength Factor* (*DPF*) is dependent on the light wavelength $\lambda$, absorption coefficient $\mu_a$, reduced scattering coefficient $\mu_s'$, and the distance from the source. This factor can be estimated by [18]:

$$DPF(\lambda,d) = \frac{1}{2}\sqrt{\frac{3\mu_s'(\lambda)}{\mu_a(\lambda)}}\left[1 - \frac{1}{1+d\sqrt{(3\mu_a(\lambda)\mu_s'(\lambda)}}\right]. \quad (1)$$

After obtaining the *DPF* value, the intensity ratio measured from the light source can be formulated as:

$$\frac{I(\lambda,d)}{I_o(\lambda)} \equiv T(d) = e^{-\mu_a(\lambda)\, d\, DPF(\lambda,d)+G(\lambda)}, \quad (2)$$

where $I_o(\lambda)$ is the light source intensity, $I(\lambda,d)$ is the light intensity at distance $d$ from the source, and $G(\lambda)$ is a medium and geometry dependent constant and largely unknown.

The required light power intensity for ChR2 signal triggering is 8-12 $mW/mm^2$ [19]. From [20], the absorption coefficient $\mu_a$ and the reduced scattering coefficient $\mu_s'$ for brain tissue are 0.07 $mm^{-1}$ and 1.404 $mm^{-1}$, respectively. Based on these values, the light intensity ratio or transmittance follows the curve depicted in Fig. 6, which decreases exponentially as the distance from the light source increases. This phenomenon occurs due to the absorption and scattering factors which are represented as the DPF value.

## V. Energy Harvesting for WiOptND

### A. Piezoelectric Nanowires

The piezoelectric material has been widely used for harvesting energy due to its unique ability to produce electric charge with respect to the applied mechanical stress. The utilization of certain piezoelectric material is based on the consideration of the type of application, power requirement, vibration frequency, and the geometry structure. Some well-known materials used are *lead zirconate titanate* (*PZT*), *aluminum nitride* (*AlN*), *barium titanate* (*BaTiO3*), and *zinc oxide* (*ZnO*) in the form of crystal or nanowires [21] [15] [22]. Taking into account the energy requirement, the WiOptND uses ZnO nanowires complemented by thin coating of (<100 $\mu$m) of acrylic *Polymethyl methacrylate*). The coating is important to avoid harmful effects on the brain tissue. Related to the power/energy conversion of the material, the electromechanical coupling coefficient is one important parameter to consider in deciding the appropriate material and geometry structure to be used in harvesting the energy [23].

The attenuation experienced by ultrasound wave depends on the frequency and the depth of the tissue. For brain tissue, the attenuation coefficient, $\alpha$, is 0.435 $dB/(cm \cdot MHz)$ [24]. The effect on the transmitted signal power intensity can be formulated as follows:

$$I_{nd} = I_s 10^{-(\alpha f d/10)}, \quad (3)$$

where $I_{nd}$ and $I_s$ are the power intensity levels at the surface of the energy harvester and the acoustic wave source, respectively, $\alpha$ is the attenuation coefficient of the brain tissue, $f$ is the ultrasound wave frequency, and $d$ is the distance between the WiOptND and the sub-dura transceiver.

According to (3), if the ultrasound source emits 720 $mW/cm^2$ wave intensity, the power for a $100 \times 100 \mu m^2$ energy harvester is ~60 $mW$. This calculation is based on the 2-$mm$ depth of the cerebral cortex. The electromechanical conversion occurs in the energy harvesting element, therefore, its conversion rate needs to be taken into consideration. Assuming that the electromechanical conversion rate, $\eta$, is

50%, the effective electrical energy generated is 30 $mW$. This result can be obtained from:

$$P_{nd} = I_{nd} A_{EH}, \tag{4}$$

$$P_e = P_{nd} \eta, \tag{5}$$

where $P_{nd}$ and $P_e$ are the power received to vibrate the nanowires of the energy harvester and the electrical power after the conversion from mechanical to electrical energy, respectively, and $A_{EH}$ is the effective surface area of the energy harvester.

### *B. Storage Capacitor*

The next stage after the energy is produced by the harvester and is rectified, is the storage of charge in the micro-supercapacitor based on the generated voltage $V_g$ from the $ZnO$ nanowires [25]. The micro-supercapacitor can be based on the interdigital electrodes of reduced graphene oxide and carbon nanotube composite [26]. This capacitor is considered the most efficient for the WiOptND due to its miniature size and large charge storage capacity. Using the power and voltage of the energy harvester, the electrical current, $i_g$, flowing in the circuit can be represented as:

$$i_g = \frac{P_e}{V_g}. \tag{6}$$

The amount of electrical charge, $\Delta Q$, supplied and stored in the storage capacitor per charge cycle, $t_{cycle}$, can be estimated based on the nanowire vibration frequency, $f$, and the current, $i_g$, flowing from the energy harvester to the storage circuit. This value can be obtained from [25]:

$$\Delta Q = i_g t_{cycle} = \frac{i_g}{f}. \tag{7}$$

### *C. Charging Cycles*

The energy from the capacitor is utilized to power the LED. For the LED, the minimum light intensity requirement should be fulfilled and at the same time having the low power demand in accordance to the power availability in the storage capacitor is important.

The time duration for charging and discharging period of the storage capacitor can be represented by the number of cycles, $n_{cycle_{charge}}$ and $n_{cycle_{discharge}}$, and this can be represented as follows [25]:

$$n_{cycle_{charge}} = \left\lceil -\frac{V_g C_{cap}}{\Delta Q} \ln\left(1 - \sqrt{\frac{2E_{max}}{C_{cap} V_g^2}}\right)\right\rceil, \tag{8a}$$

$$n_{cycle_{discharge}} = \left\lceil -\frac{V_g C_{cap}}{\Delta Q} \ln\left(\sqrt{\frac{2E_{max}}{C_{cap} V_g^2}}\right)\right\rceil, \tag{8b}$$

where $E_{max}$ is the maximum electrical energy that can be stored in the capacitor, $C_{cap}$.

The voltage level in every cycle can also be determined for both the charging and discharge processes. The instantaneous voltage level in terms of cycle numbers is represented as:

$$V_{cap_{charge}}(n_{cycle}) = V_g \left(1 - e^{-\frac{n_{cycle}\Delta Q}{V_g C_{cap}}}\right), \tag{9a}$$

$$V_{cap_{discharge}}(n_{cycle}) = V_g e^{-\frac{n_{cycle}\Delta Q}{V_g C_{cap}}}. \tag{9b}$$

TABLE I: Simulation Parameters

| Parameter | Value [Unit] | Description |
|---|---|---|
| WiOptND density | 0.024 to 1.2 [$/cm^3$] | Randomly scattered |
| Ultrasound Frequency | 500 to 3M [$Hz$] | - |
| Cortical Cortex Depth | 2 to 4 [$mm$] | - |
| Neural spike period | $\lambda$ = 6 [$ms$] | Exponential dist. |
| Data sample | 10,000 | Randomly generated |
| Nanowire surface area | $10^4$ to $2\times10^4$ [$\mu m^2$] | Energy harvester |

### *D. WiOptND Energy and Power Evaluation*

In this section, we numerically evaluate the energy and power storage circuitry of the WiOptND. The parameters used for the simulations are presented in Table I. Since the duration of the charging and discharging of storage capacitor is sufficiently fast regardless of the 2-$mm$ thickness of the cerebral cortex layer and operating frequencies variants, the main concern lies in the electrical specifications of the light source component and energy harvester. This, in turn, affects the constant intrinsic values of the storage capacitor.

Our analysis is based on determining the radiated intensity from the LED in order to obtain desired light intensity on the target neuron. Fig. 7(a) shows the result of the required emitted light intensity of the LED to achieve the level inside the range of optogenetics stimulating intensity, which is 8-12 $mW/mm^2$ with respect to distance. Similar to the previous calculations, Fig. 7(b) and 7(c) illustrates the effect of the storage capacitor component to the required light intensity by the optogenetics. For both charging and discharging processes, higher light intensity exhibits faster periods related to larger $\Delta Q$ electrical charge. Fig. 7(d) and 7(e) presents the difference of energy storage phenomena when the effective area of the energy harvester is doubled. Larger effective surface area leads to higher electrical charge supply. However, when the frequency is varied, no significant change is noticeable in term of energy in the storage capacitor, which is depicted in Fig. 7(f).

The relation between the charging time duration and the piezoelectric nanowires surface area with varying ultrasound frequencies is analyzed in Fig. 8. From this simulation, it can be observed that the surface area of the energy harvester is linearly proportional to the generated energy resulting in faster charging process. As shown in the result, the differences in the frequencies have no effect on the quantity of stored energy.

## VI. SYSTEM CHARGING PROTOCOLS

While the previous sections discussed the functionalities as well as system performance of each device, this section will discuss its operation as a network. Fig. 9 illustrates the heterogeneous nature and density variation of neuronal networks in a cortical cortex. The deployment and topology of the WiOptNDs are highly correlated to the network structure and characteristics of the neuronal networks. To represent the basic system architecture, three element sets should be defined, i.e., $L = \{2/3, 4, 5, 6\}$ which represents the set of cortical layers, $F = \{f_1, f_2, f_3, ..., f_n\}$ which represents the list of transmitted frequencies, and ND = {$ND_1$, $ND_2$, $ND_3$, ... $ND_m$}, $\{n, m\} \in \mathbb{N}^*_{\leq n}$ is for the set of WiOptNDs signalled by the sub-dura transceiver, $Tx_{sub}$. Additionally, $Tx_{sub}(F)$ represents the transmitting frequency $F$ emitted by the sub-dura transceiver. Since multiple devices are concerned, and energy harvesting is required from the sub-dura transceiver, a charging protocol is required. The charging protocols that will be discussed in this section range from a simple *Charge and Fire*, to more complex protocols that will maximize energy efficiency, which include *Predictive Sliding Detection Window* and its variant the *Markov-Chain based Time-Delay Patterns* protocols.

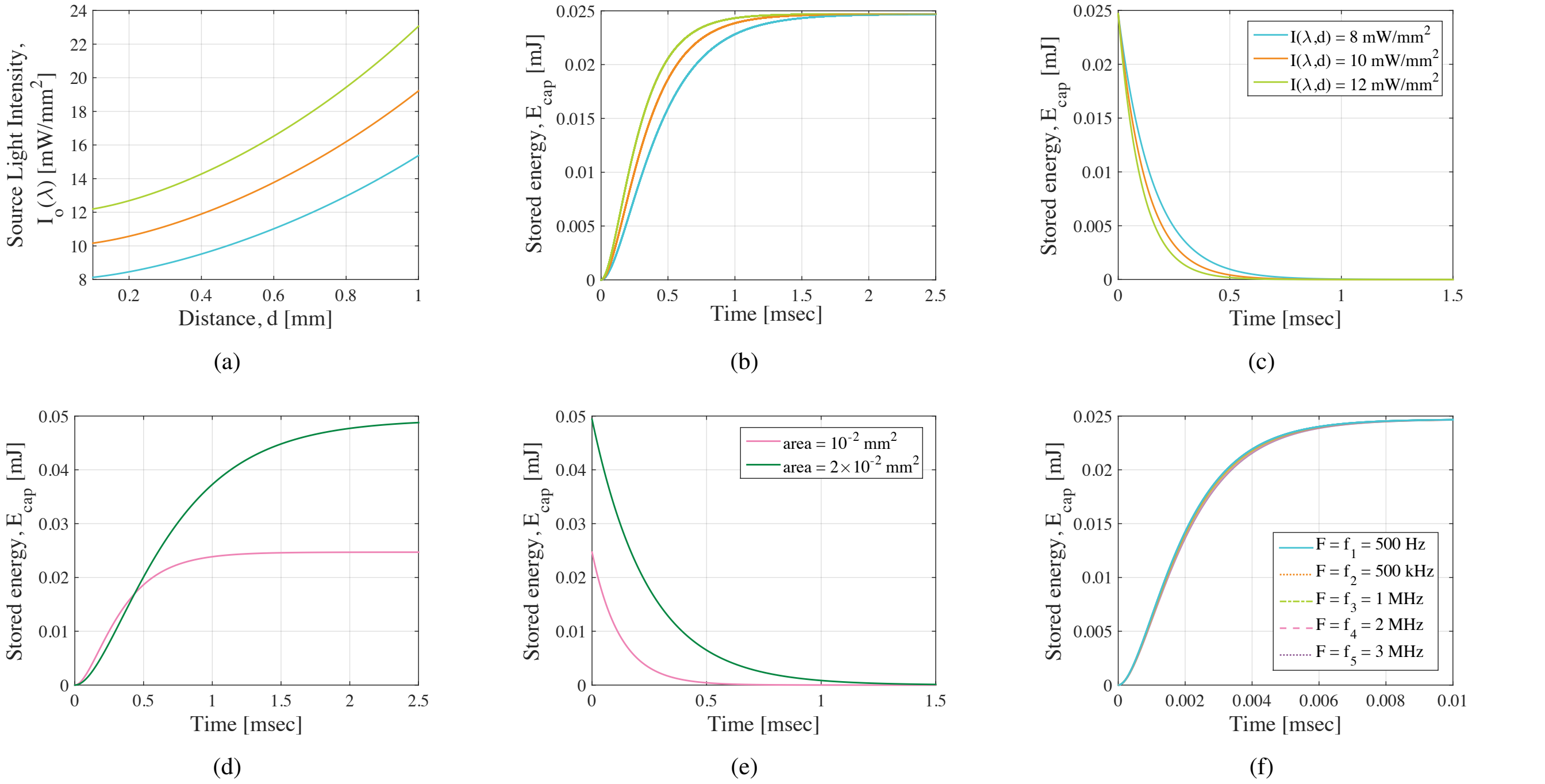

Fig. 7: (a) Intensity at the light source as a function of distance with variations in the required intensity for the optogenetics. (b) Illustration of storage energy during the charging period as a function of time with light intensity variations, and a constant frequency of 500 $Hz$ at 500$\mu m$ distance. (c) Illustration of storage energy during the discharging period as a function of time with light intensity variations, and a constant frequency of 500 $Hz$ at 500$\mu m$ distance. (d) Illustration of storage energy during the charging period as a function of time with energy harvester effective area variations, and $I(\lambda, d)$ of 10$mW/mm^2$ with constant frequency of 500 $Hz$ at 500$\mu m$ distance. (e) Illustration of storage energy during the discharging period as a function of time with energy harvester nanowire area variations, and $I(\lambda, d)$ of 10$mW/mm^2$ with constant frequency of 500 $Hz$ at 500$\mu m$ distance. (f) Illustration of storage energy charging process with ultrasound frequency variations.

### A. *Charge and Fire Protocol*

For this protocol implementation, the sub-dura transceiver transmits one frequency, $f_i \in F$, which corresponds to one specific WiOptND when signaling to stimulate the neuron is required. Considering $s(L[n], t)$ is the firing state of a neuron of $n^{th}$ layer in time slot $t$, $s \in \{0, 1\}$, the frequency transmission process could be translated as $s[L[y], t] = 1 \rightarrow Tx_{sub}\ (f_n)$. The protocol operating principle is as follows: The full neuron firing sequence raster plot knowledge is held inside the sub-dura transceiver. The sub-dura transceiver also has the knowledge of which WiOptND device is interfaced to a specific neuron. Based on a time-division access scheme, as the sub-dura transceiver scans through the raster plot and encounters a signal that needs to be stimulated, it emits an ultrasound frequency $f_i$, which charges the device. The design of the circuitry is very simple, as soon as the ultrasound frequency is emitted, it immediately charges to the full capacity of the micro-supercapacitor, $E[n, t] \rightarrow E_{max}$, which in turn discharges and powers the LED. This leads to the illumination of the neuron. Based on our analysis in Fig. 7(b), the ultrasound wave emission will last approximately 1 $ms$ in order to fully charge the micro-supercapacitor to maximum capacity.

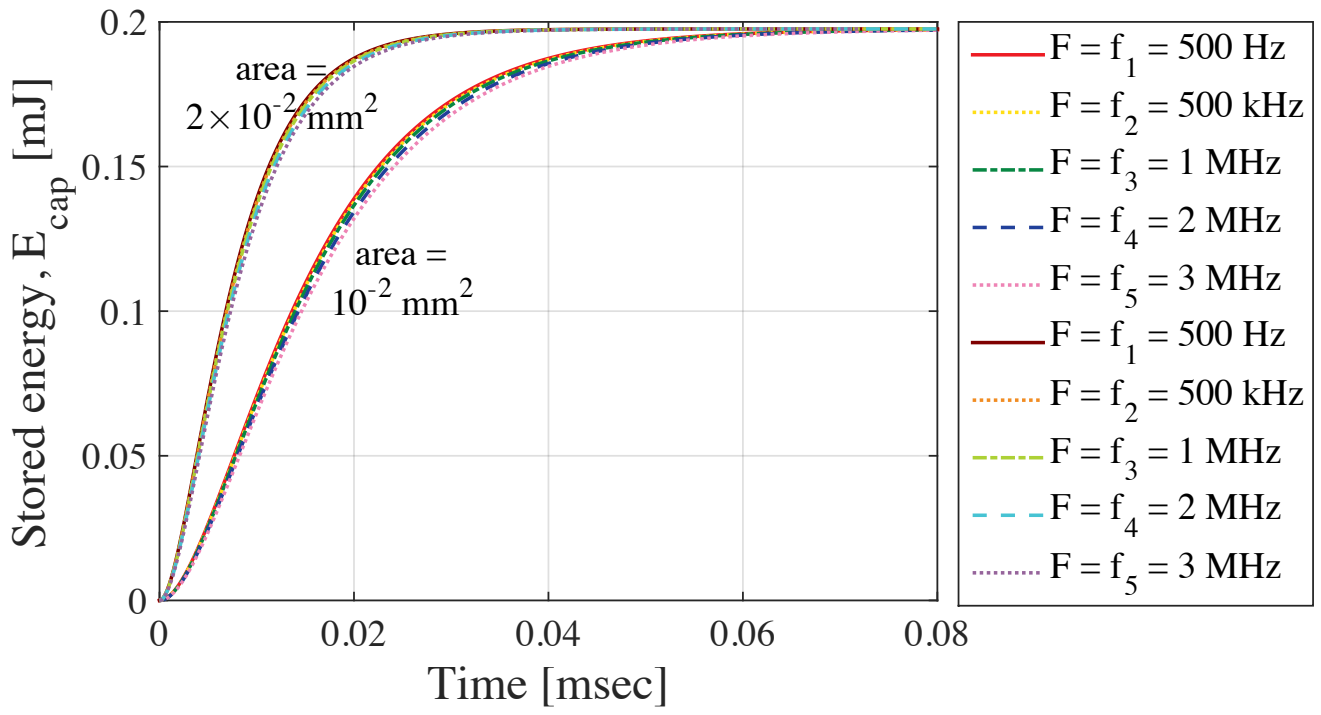

Fig. 8: Illustration of storage energy performance as a function of time with ultrasound frequency variations. The performance compares the storage energy performance for two different nanowires surface area.

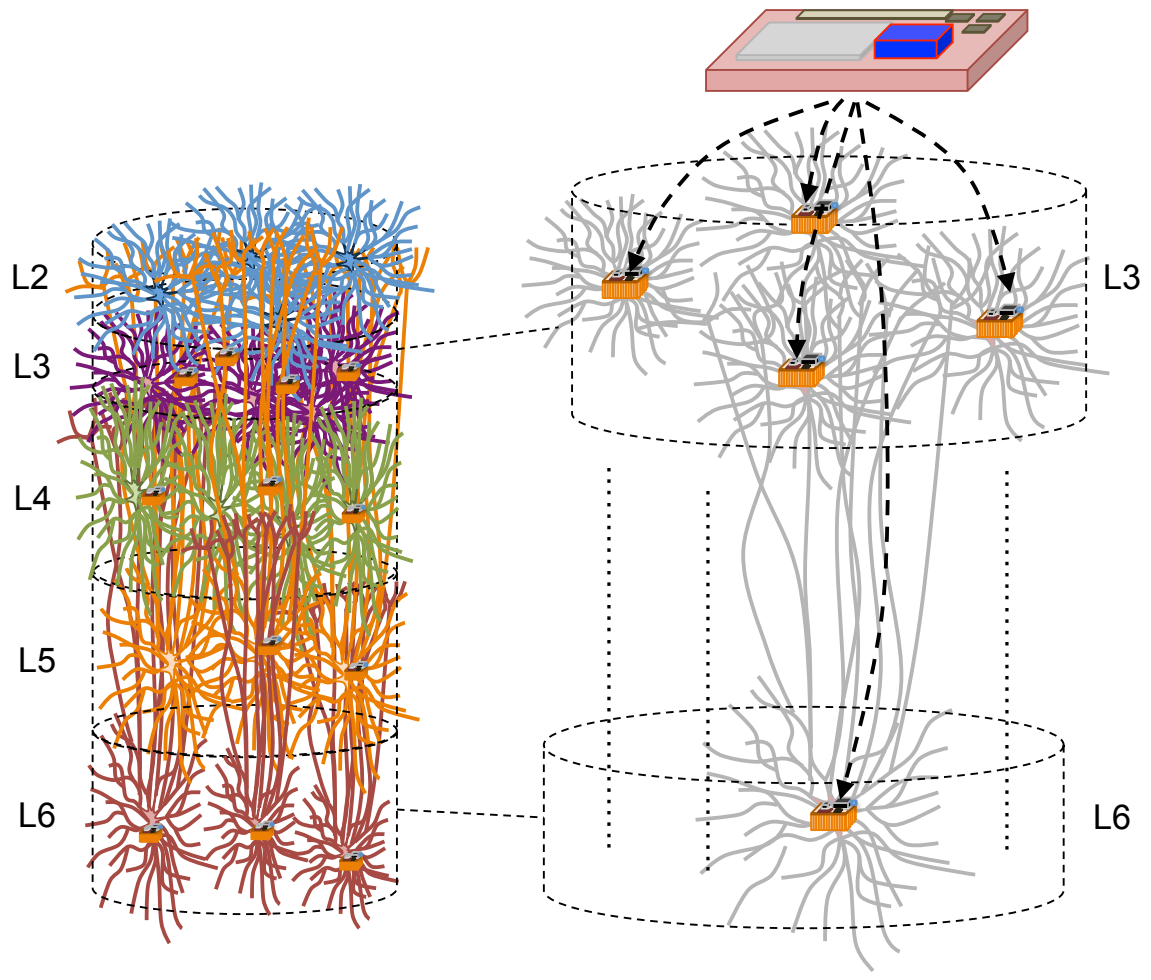

Fig. 9: WiOptND nanonetworks deployed in the cortical column, between $L2/3$ - $L6$.

As an example, Fig. 10 illustrates how this protocol handles the firing pattern for three WiOptNDs. As shown in the figure, each of the WiOptND has a unique ultrasound frequency, which is based on the specific resonant frequency of the piezo element of the VOX. At approximately $t_5$, a misfiring occurs for WiOptND$_3$ and this is due to the clash in time-slot with WiOptND$_1$ (in our proposed approach, the sub-dura transceiver can only emit ultrasound with a single frequency). While the protocol is very simple, and requires very basic circuitry, the major drawback is that the sub-dura transceiver is required to emit ultrasound signals at various frequencies. This also becomes a major challenge, when we consider that the piezo element for addressing will also require different resonant frequencies that have considerable spacings to not lead to overlap with signaling other devices. Another limitation is that the sub-dura transceiver is required to have full knowledge of the raster plots for all the cortical circuit functionalities.

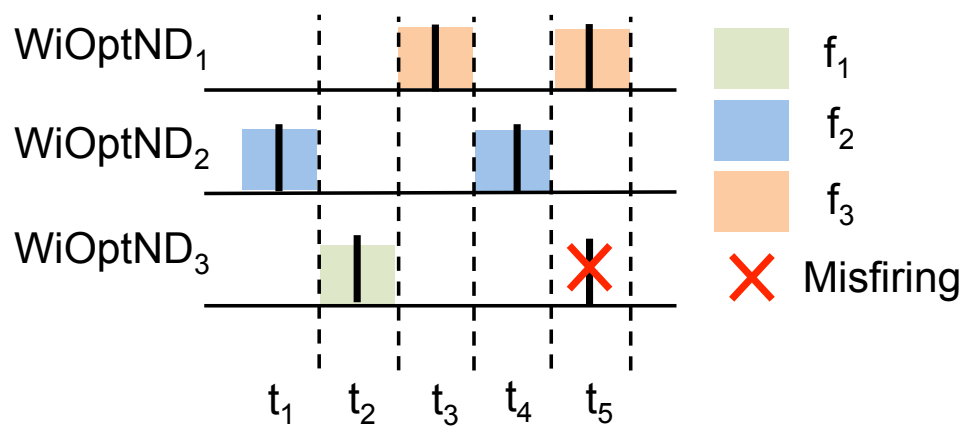

Fig. 10: *Charge and Fire* employs one-to-one relation between frequency transmission from the sub-dura transceiver, and the neural spike sequence. More than one neural spikes in a single time-slot leads to a misfiring event.

## B. Predictive Sliding Detection Window

Since ultrasound signals will emanate to the entire region of the nanonetwork, this also means that all devices will automatically get charged each time a sub-dura transceiver emits an ultrasound signal. However, the fact that each device requires a unique resonant frequency addressing scheme, means that in the *Charge and Fire* protocol, the sub-dura transceiver will still need to signal each device. This results in excessive energy depletion of the sub-dura transceiver, and waste of repeating charging signals that are emitted to devices that have already been charged. At the same time, the diverse charging frequencies may be limited by the piezo element technologies that is used for the addressing scheme. Therefore, to address this issue, this section presents the *Predictive Sliding Detection Window*, which aims to, (1) lower the number of frequencies that are required to signal to all the WiOptND devices, and (2) minimize the emitted ultrasound frequencies from the sub-dura transceiver to lower the energy depletion, by exploiting a parallel charging scheme. In the case of (1), the different devices should respond to the same ultrasound frequency to charge multiple devices.

*1) Parallel Charging:* Fig. 11 illustrates an example of the *Predictive Sliding Detection Window* charging protocol. In this example, there are three different WiOptND devices and three ultrasound charging frequencies. As illustrated in Fig. 11 (a), each of the frequencies and the WiOptND devices forms a matrix $M_{ii}$ that represents a random simulated pattern with respect to time. For example, ultrasound frequency $f_1$ signals WiOptND$_2$ in the first time-slot, $t = t_0$, followed by WiOptND$_3$ after one time-slot delay, $t = t_0 + 1$, and finally WiOptND$_1$ after another time-slot delay, $t = t_0 + 2$. This means that if ultrasound frequency $f_1$ is emitted at $t_0$, there needs to be a time delay for WiOptND$_3$ and WiOptND$_1$ before they can discharge to light the LED and stimulate the neuron. Frequencies $f_2$ and $f_3$ utilize similar three time-slot durations but implementing different pattern predictions as illustrated in Fig. 11 (a). The total frequencies and time-slots of one pattern can be dubbed as the size of the sliding detection window.

In Fig. 11 (b) - (f), the sliding detection window protocol is illustrated. The objective here is to slide the detection window and find the overlaps between the neuron firing sequence and time-slot of the pattern prediction for discharging the device. Starting at $t_1$ in Fig. 11 (b), the pattern prediction sequence for frequency $f_1$ perfectly matches to the sequence of spikes for WiOptND$_1$ and WiOptND$_3$, and compared to other frequencies, this window is able to parallel charge the highest number of devices. Due to this reason, $f_1$ is the selected frequency for this window. As the sliding window moves along, the time-delay times are matched to the spikes for each time-slot to decide which charging frequency will be emitted next. Fig. 11 (c) shows that for all the three frequencies, only one pattern matches to a device with a neural spike, and that this is WiOptND$_2$. However, moving the sliding detection window along in Fig. 11 (d), $f_2$ ultrasound charging pattern overlaps with two neural spikes of WiOptND$_1$ and WiOptND$_3$. The remaining process is illustrated for Fig. 11 (e) and (f).

Algorithm 1 explains in more detail the *Predictive Sliding*

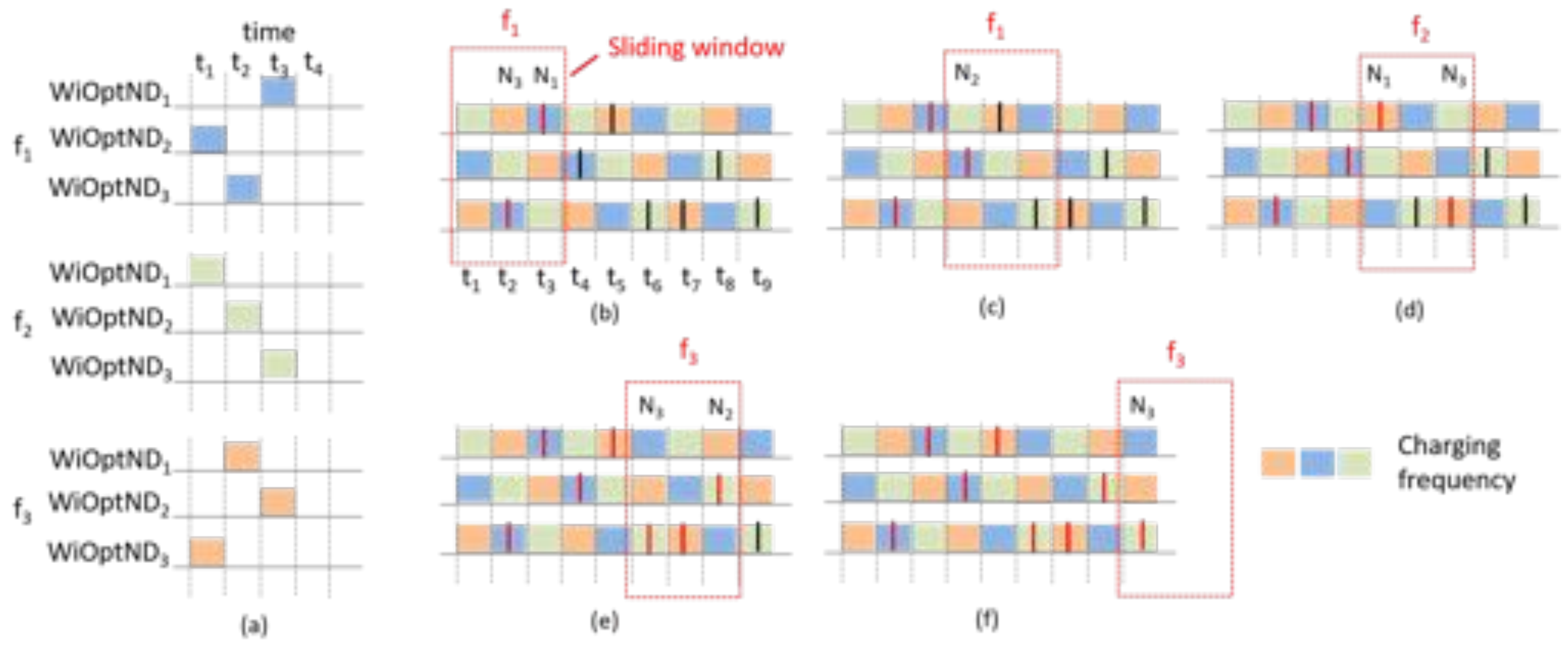

Fig. 11: Illustration of Sliding Detection Window mechanism utilizing three different frequencies/patterns. (a) presents the predicted patterns for 3 different frequencies for 3 WiOptNDs with respect to time. (b) - (f) illustrates the sliding window with respect to time, and the frequency used to charge the devices. The selected frequency is based on the highest number of parallel devices that can be charged in that window.

*Detection Window* protocol. Initialization of the number of patterns and time-delays can be represented as the array matrix $P[1..n]$ where $n$ is the total amount of neural spiking patterns. Based on the pattern prediction process, matrix $M_{ij}$ is obtained which will be compared with the known neural spikes within each window. Before moving to the next phase, this protocol makes sure if the first time-slot of the detection window contains at least a spike. Afterwards, matrix $M_{ij}$ will be truncated so that the size is matched with the matrix of the patterns. This truncated matrix $M_{ii}$ is compared against all the patterns and the highest number of overlapping pattern is selected by emitting the corresponding frequency. Finally, when the selected frequency $f_i$ has been radiated, corresponding spikes will be omitted and the protocol will analyze the next time-slot until it reaches the end of the pattern sequence.

**Algorithm 1** Predictive Sliding Detection Window

```
Initialize P[1...n]   ▷ where n is total number of available frequencies
M_ij ← PatternPrediction
for a = 1 to j do
    M_ii ← M[:, a : a − 1 + column(P[b])]
    if M[:, 1] ≠ 0  &  M_ii ≠ 0 then
        for b = 1 to n do
            simTest ← compare M_ii == P[b]
        end for
        maxSim ← max(simTest)
        tempFiringSlot ← 2 × P[maxSim] − M_ii
        M_ii ← tempFiringSlot
    end if
end for
```

*2) Circuitry:* The challenge is that since each device is interfaced to different neurons that spike at different time-slots, the charging process must consider the timing difference between the spikes when a single charging ultrasound frequency is emitted. This means that when an ultrasound frequency is emitted at $t_{0+j}$, a time-delay for a predicted overlap at $t > t_{0+j}$ is required for the circuit. This time delay will count down until the specific time-slot has arrived, at which point the charge will be released from the capacitor to light the LED. This could be achieved by adding a time-delay circuitry that extends over the original circuit presented in Fig. 5, and used in the *Charge and Fire* protocol. The circuit design of the WiOptND for the *Predictive Sliding Detection Window* is illustrated in Fig. 12. In this circuit, the time-delay relay circuit is situated between the storage capacitor and the LED. The number of frequency-dependent switch and time-delay relay pairs corresponds to the amount of operating frequencies used in the system. For instance, the set of frequencies $F = \{f_1, f_2, f_3, ..., f_i\}$, where the complete pattern that can be formed using $i$ number of frequencies is $i!$ (factorial of $i$). However, the designated number of patterns are kept as an independent variable during the design process.

*3) Neural Spike Prediction:* One limitation with the *Charge and Fire* protocol, is that the sub-dura transceiver is required to have the full knowledge of the raster plot neural spike sequence. This could be a major challenge, given that the sequence changes for variations in activities (this will be illustrated later in Sec. VI-B4, when we demonstrate the changes in raster plot of visual cortex of experiments conducted on the *macaque monkeys*). Since the *Predictive Sliding Detection Window* protocol is already predicting the neural spikes, one approach is to augment the protocol with existing neural spike prediction solutions. In particular, since the sliding window is required to scan future spikes, a prediction process can be integrated. Numerous research has investigated prediction processes for neural spikes, where in majority of the cases *in-vivo* neuronal system have been known to contain patterns that corresponds to a certain stimulus [27]. For example, when *retinal cells* receive visual information, this data is conveyed through the optic nerve and stimulates neurons firing in the *V1 primary visual cortex*. The pattern of the neural spikes is directly related to the light intensity, and determines four related parameters which includes the time of occurrence, number of spikes, jitter periods, and number of jitters [28]. In this paper, we employ the neural spike prediction process which is illustrated in Fig. 13). In this scenario, a light to electrical converter will wirelessly transmit signals to the external transceiver based on changes in the light intensity, which is then transmitted to the sub-dura transceiver, and this is used to guide the charging protocol.

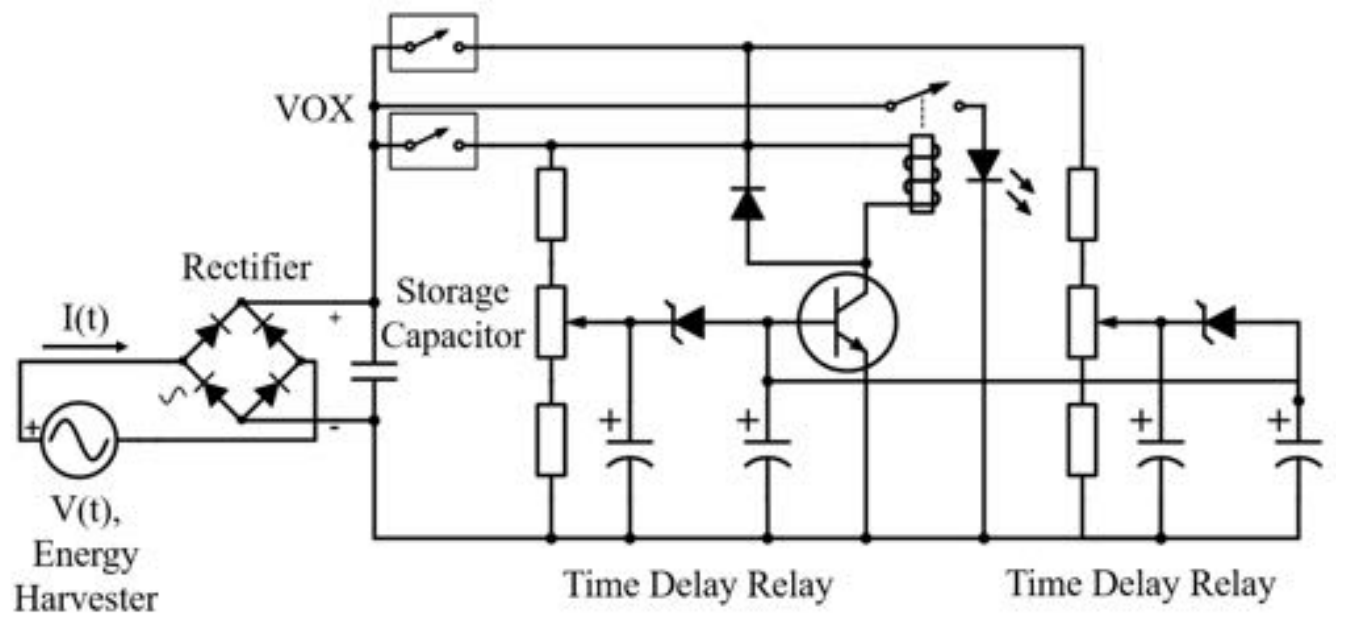

Fig. 12: WiOptND circuit diagram with integrated time-delay relay for the *Predictive Sliding Detection Window* protocol and its variant the *Markov-Chain based Time-Delay Patterns* protocol.

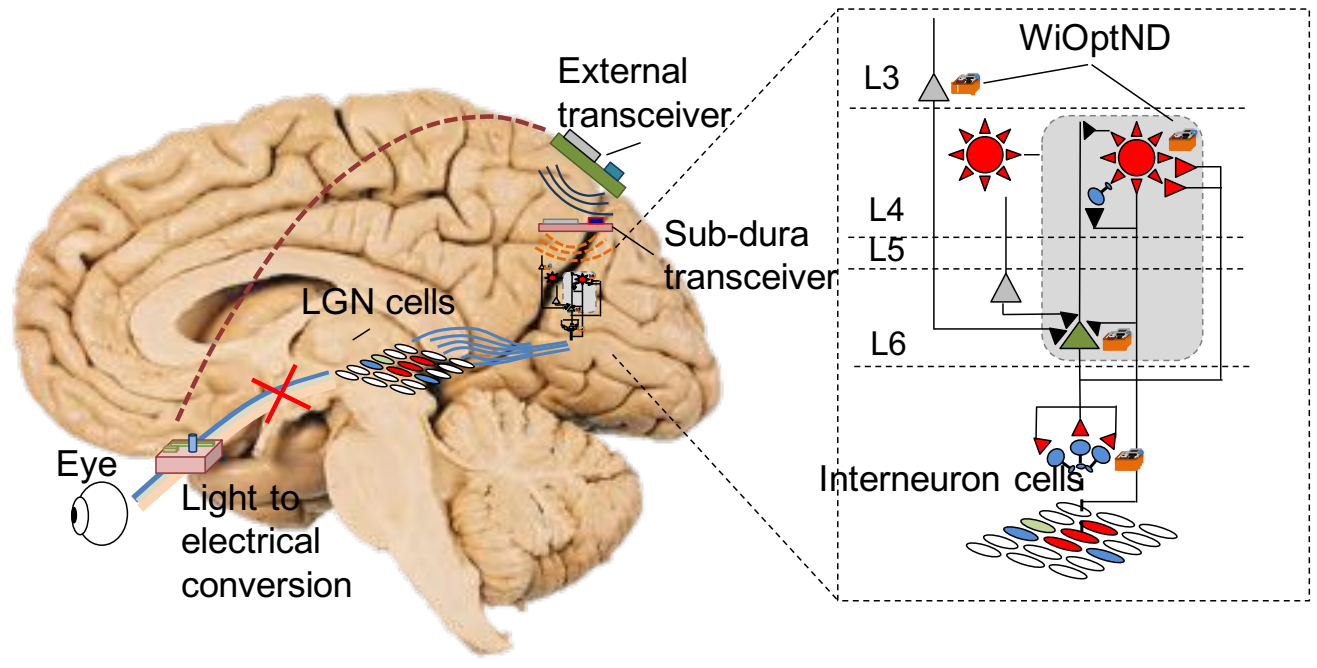

Fig. 13: Example deployment of the WiOptND nanonetwork in the Brain Visual Cortex. The neural circuit connection to the $V1$ primary visual cortex is impaired, requiring the deployment of WiOptNDs, where its coordinated stimulation will compensate for the failed neurons.

Based on the mathematical model in [28], the pattern prediction maps of the firing rate and spiking sequence is formulated as follows:

$$r(t) = \delta(h(t)) - \theta)\dot{h}(t)H(\dot{h}(t)) \tag{10}$$

where

$$h(t) = g(t) + a(t) + \int_{-\infty}^{t} r(\tau)(1+b(\tau))P(t-\tau)\,d\tau \tag{11}$$

$$g(t) = \int_{-\infty}^{t} s(\tau)F(t-\tau)\,d\tau \tag{12}$$

$$H(x) = \begin{cases} 1, & \text{if } x > 0 \\ 0, & \text{otherwise.} \end{cases} \tag{13}$$

where $r(t)$ is the sum of delta function spikes at one particular time instance when $h(t)$ crosses the threshold $\theta$ with a positive gradient slope. The function $g(t)$ defines the filtered time domain response for stimulus $s(t)$ and is obtained from the convolution operation with filter $F(t)$. The parameters $a(t)$ and $b(t)$ represent Gaussian noises, while $P(t)$ is a feedback potential. This particular model has proved to perform well when compared with real neural spiking patterns.

TABLE II: Simulation Parameters

| Parameter | Value [Unit] | Description |
|---|---|---|
| Neural spike rate | 100:5:130 $[Hz]$ | Exponential dist. |
| Cortical column layers | 4 $[layers]$ | - |
| Number of predicted patterns | 5, 10, 20 | No. of Ultra. freq. |
| Raster plot period | 10 $[sec]$ | - |
| Number of data | 10 $[cycles]$ | Sim. cycles/freq. |

*4) Evaluation:* In this section, we evaluate the performance of the *Predictive Sliding Detection Window* protocol, and compare this to the *Charge and Fire* protocol, using the light stimulated neural spike prediction model of [28] as a case study. We will first define the metrics that is used in our evaluation. The total neural spike misfiring number, $n_{mis}$, is represented as:

$$n_{mis} = \sum_{t=0}^{T}\sum_{y=0}^{|L|}\left[\sum_{k=0}^{|L|}[\min(s(L[y],t), s(L[k],t))\right] + H[L[y],t] \tag{14}$$

where

$$H[n,t] = \begin{cases} 0, & \text{if } E[n,t] < E_{max}, \\ 1, & \text{if } E[n,t] = E_{max}. \end{cases} \tag{15}$$

which represents whether the stored energy $E[n,t]$ is sufficient to turn on the LED at time $t$, by comparing it to the required energy $E_{max}$.

The neural spike misfiring ratio, stimulation efficiency ratio, and stimulation ratio can be formulated as:

$$\gamma_{mis}(L[y]) = \frac{n_{mis}}{\sum_{t=0}^{T} s[L[y],t]} \tag{16}$$

$$\eta_{stim}(L[y]) = 100\% - \gamma_{mis}(L[y]) \tag{17}$$

$$\gamma_{stim}(L[y]) = \frac{\sum_{t=0}^{T} |Tx_{sub}[f_n,t]|}{\sum_{t=0}^{T} s[L[y],t]} \tag{18}$$

where $\sum_{t=0}^{T} |Tx_{sub}[f_n,t]|$ is the total number of frequency transmission during period of $T$.

As mentioned above, the time-delay patterns affect the matching probability to the predicted neural spike sequence and this is related to the number of ultrasound resonant frequencies. In order to evaluate the protocols, simulations were conducted in Matlab with the parameters presented in Table II. The results presented in Fig. 14(a) and 14(b) compare the performance of the *Charge and Fire* and *Predictive Sliding Detection Window* protocols with respect to variations in the spike frequencies.

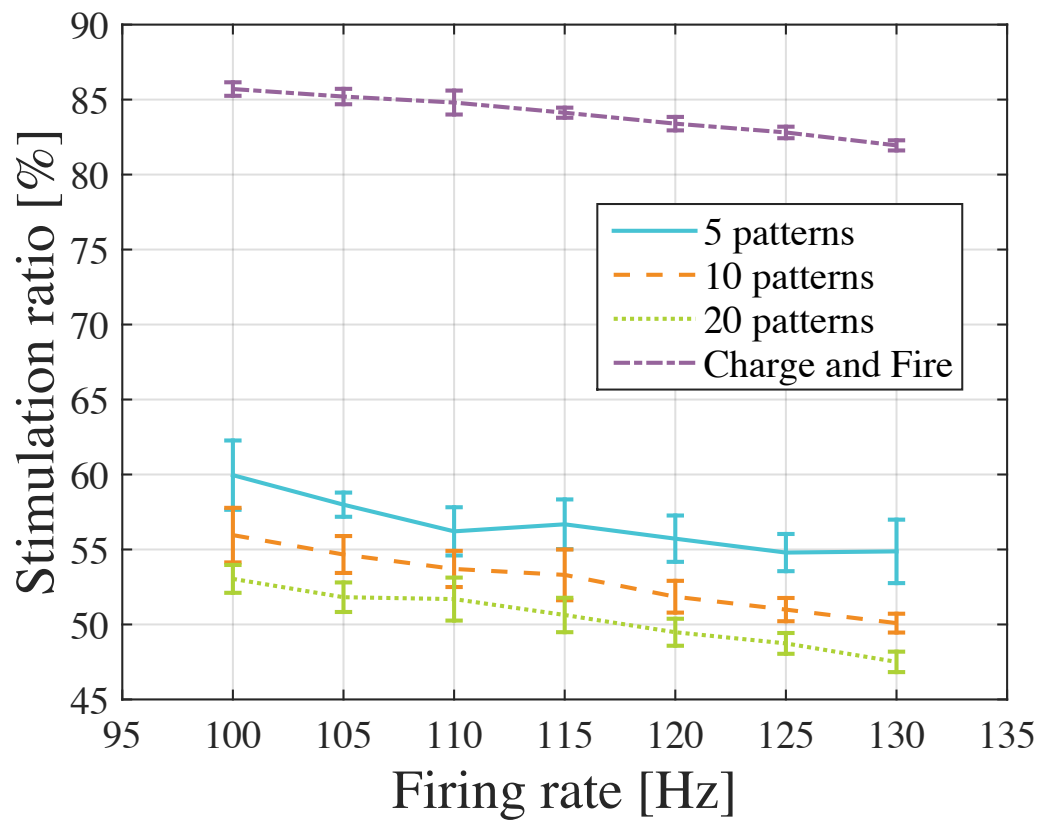

(a) Stimulation ratio vs Spike frequency.

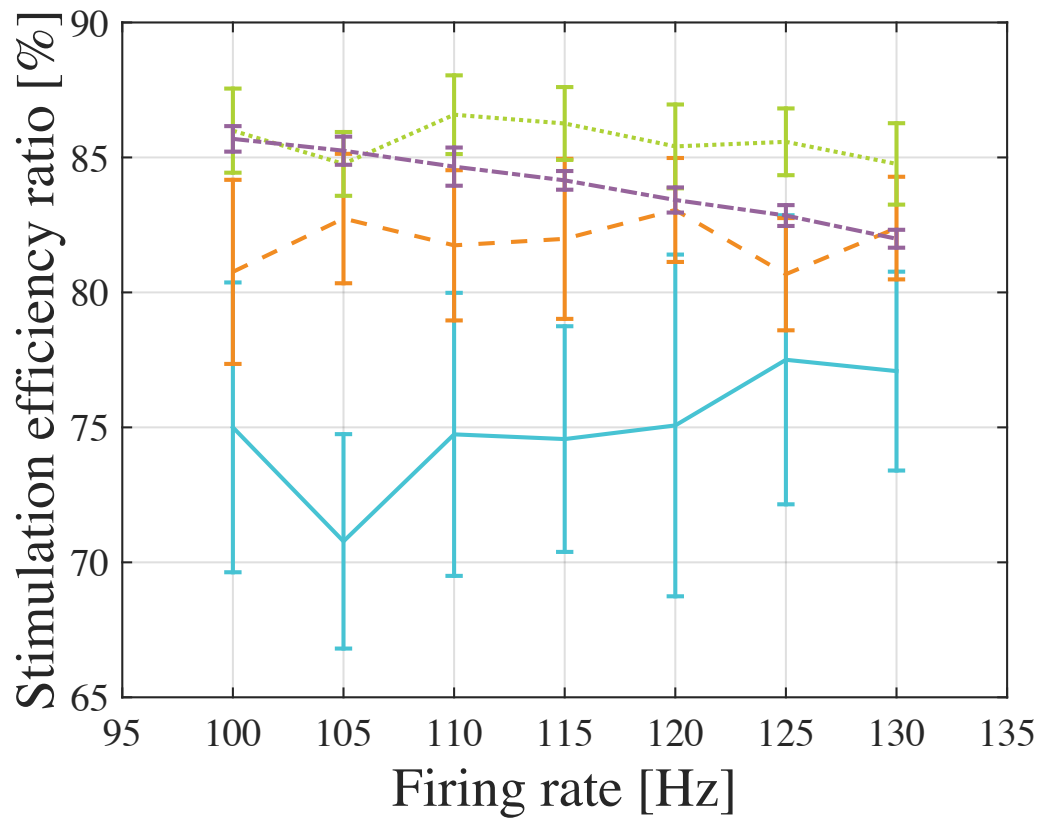

(b) Stimulation efficiency ratio vs Spike frequency.

Fig. 14: Comparison of the stimulation and efficiency ratio for the randomly chosen patterns of the *Predictive Sliding Detection Window* and *Charge and Fire* protocols. The simulation generates the average and standard deviation values for WiOptNDs deployed in four layers of the cortical column of the cerebral cortex.

The results show that improved performance results from higher neural firing rate as the stimulation ratio decreases. Stimulation ratio (18) defines how many times the sub-dura transceiver transmits the frequencies in comparison to the total amount of neural spikes. However, in terms of stimulation efficiency ratio (17) which defines the total amount of spikes successfully targeted during the stimulation process compared to the total amount of spikes, the randomness in the results is observed, and this can be attributed to the randomly chosen pattern predictions used.

Fig. 15(a) and 15(b) presents the performance analysis when the number of ultrasound frequencies/patterns and number of devices are varied. The number of devices has significant effect on the stimulation ratio, while the number of frequencies/patterns does not make a significant impact. This is due to the increased possibility of targeting multiple neural spikes based on using a smaller number of frequency transmission.

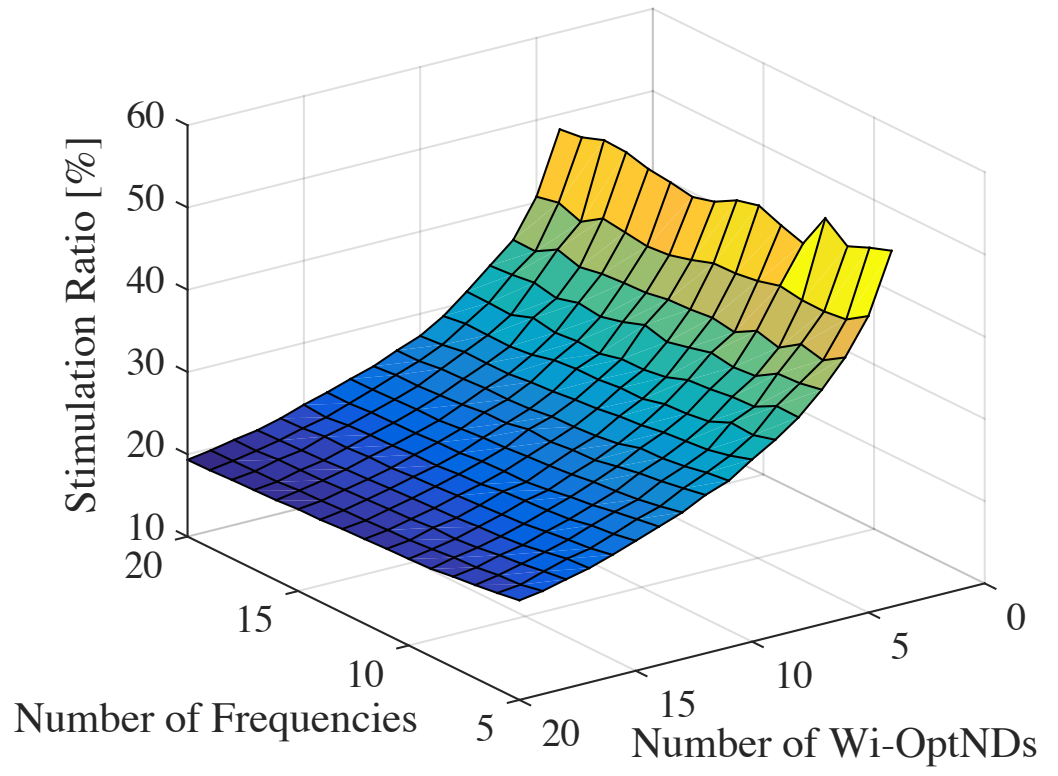

(a)

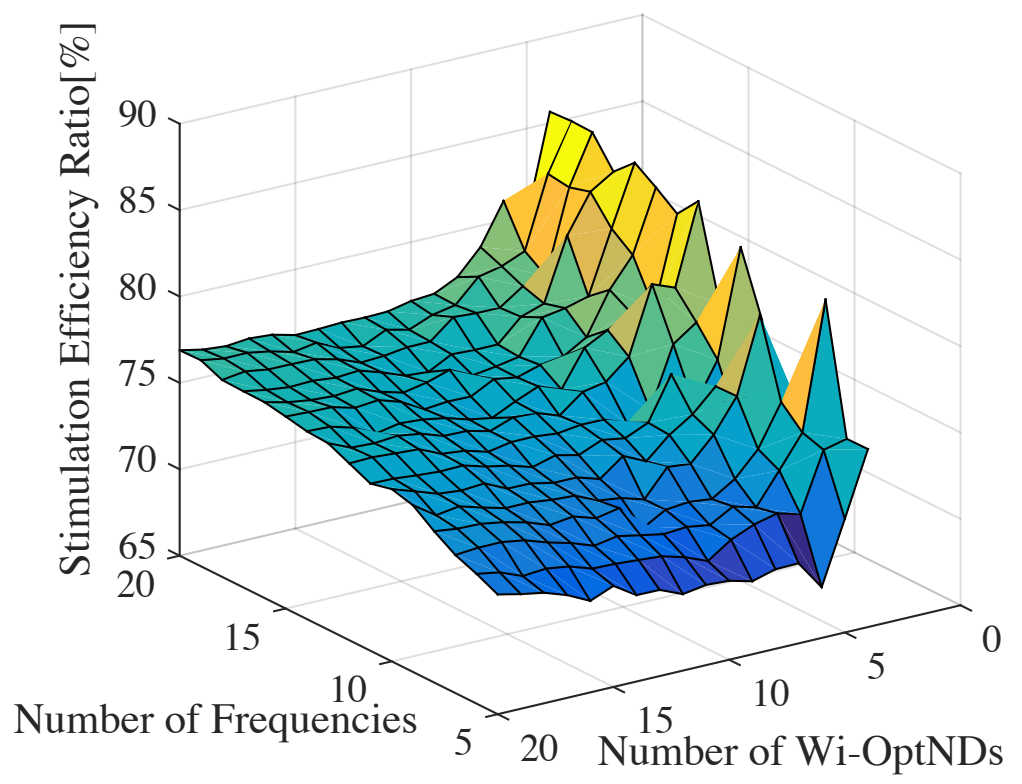

(b)

Fig. 15: The effect of the radiated ultrasound frequency quantity and the number of WiOptNDs on stimulation and efficiency ratio for *Predictive Sliding Detection Window*. Utilizing higher number of ultrasound frequencies/patterns does not have significant effect on the stimulation ratio, but it results in each stimulation process to be more efficient by targeting the desired neurons.

Our evaluation also considers how the changes in the firing patterns can affect the misfiring ratio. In this evaluation, we utilized the raster plot of the *Middle Temporal Cortex* neurons of a *macaque monkeys* when the visual image is dynamically changed. Fig. 16(a) illustrates two disks with moving dots that were used in the experiments. The initial image is presented on the left, where dots from Direction 1 and 2 are slowly moving from the center to the circle perimeter. This movement is later shifted as shown in the disk on the right. As we can see from the raster plot in Fig. 16(b), this small change in the image can totally change the sequence of neural spike patterns.

This raster plot is related to the visual stimulus $s(t)$, which based on the model in (10) will yield to the predicted sequence $r(t)$ [29]. Tuning curve represents the graphical presentation of the neurons as a result of changes in the stimuli. For example, the tuning curve can provide firing rate fluctuation information as the angle of stimulus is varied (Fig. 16(a)). As the spike frequency increases, both stimulation and misfiring events are more likely to occur. This is simulated in Fig. 16(c) and 16(d) for both the *Predictive Sliding Detection Window* and the *Charge and Fire* protocols. However, we can observe that for lower spike frequency, the *Charge and Fire* protocol experiences less misfirings since it scans each time-slots one by one, unlike the *Predictive Sliding Detection Window*, which uses the time-slots pattern matching based on the size of the window. The lower stimulation number in the plots also indicates the smaller number of radiated ultrasound signals from the sub-dura transceiver.

### C. Markov-Chain based Time-Delay Patterns

While the previous section demonstrated the benefits of the *Predictive Sliding Detection Window* protocol, one issue is the generation of the pattern prediction sequences, which was randomly generated. In order to improve the accuracy, knowledge of the connectivity in the cortical layers could be utilized to determine the firing order patterns between neurons of each layer. This could be used to minimize the inaccuracies between the pattern prediction and the target neural spikes, which in turn minimizes the energy expenditure from the sub-dura transceiver. This section will discuss how the knowledge connectivity of the different neurons in the cortical layers can be utilized to improve the time-delay patterns.

Cortical columns of the brain gray matter are characterized by highly sophisticated connections for both the intra and inter layers. This complexity is largely based on the large number of neurons with over 125 trillions of synapses in the cortex alone [30]. There is an immeasurable effort from the neuroscience community in modelling cortex connections, and one proposal is a *discrete-time* Markov chain with $|L|$ states, each representing one layer of the cortical column. The transition probability matrix $Pr$ is characterized by $|L| \times |L|$ elements, and $P$ should satisfy $\forall i, j, Pr_{i,j} \in [0, 1]$, and, $\forall i, \sum_{j=1}^{|L|} Pr_{i,j} = 1$. The markov chain representing the connectivity between the six layers of the cortical column is represented in Fig. 17 [31].

TABLE III: Connection flow probability among cortical layers.

| | | Postsynaptic neuron | | | |
|---|---|---|---|---|---|
| | | II/III | IV | V | VI |
| Presynaptic | II/III | | 0.2 | 0.27 | 0.055 |
| | IV | 0.25 | | 0.325 | 0.095 |
| | V | 0.175 | 0.15 | | 0.325 |
| | VI | 0.055 | 0.2 | 0.225 | |

Even though synapses can be stimulated from different layers of the cortical column containing the WiOptND nanonetworks, the misfiring of neural spikes may still occur. This scenario can result from the frequent number of optogenetic

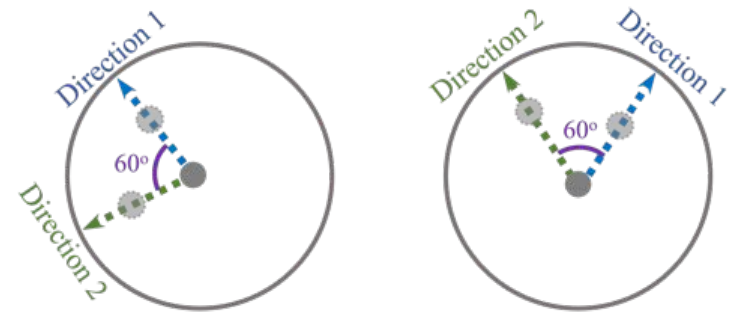

(a)

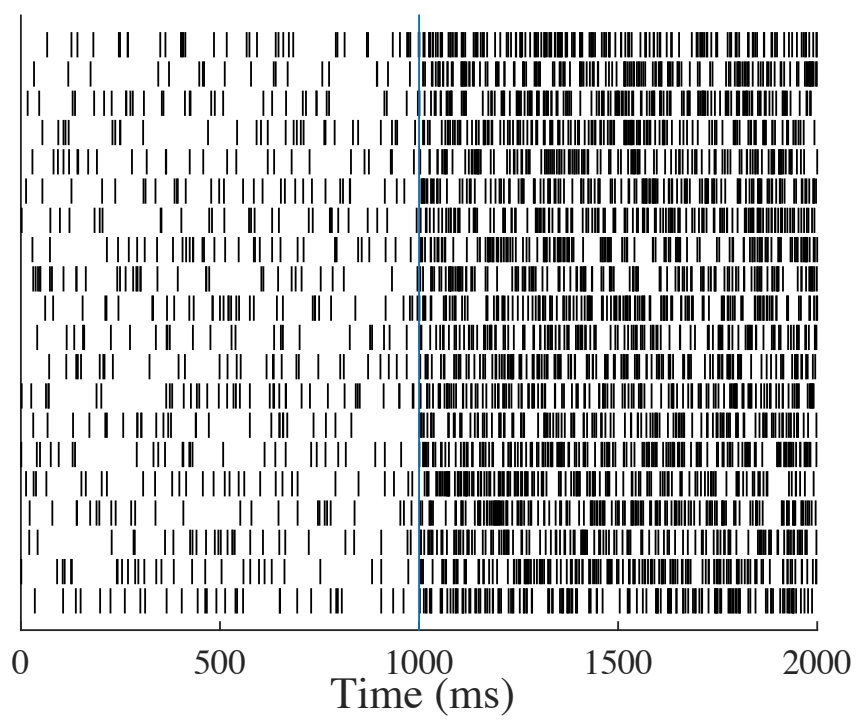

(b)

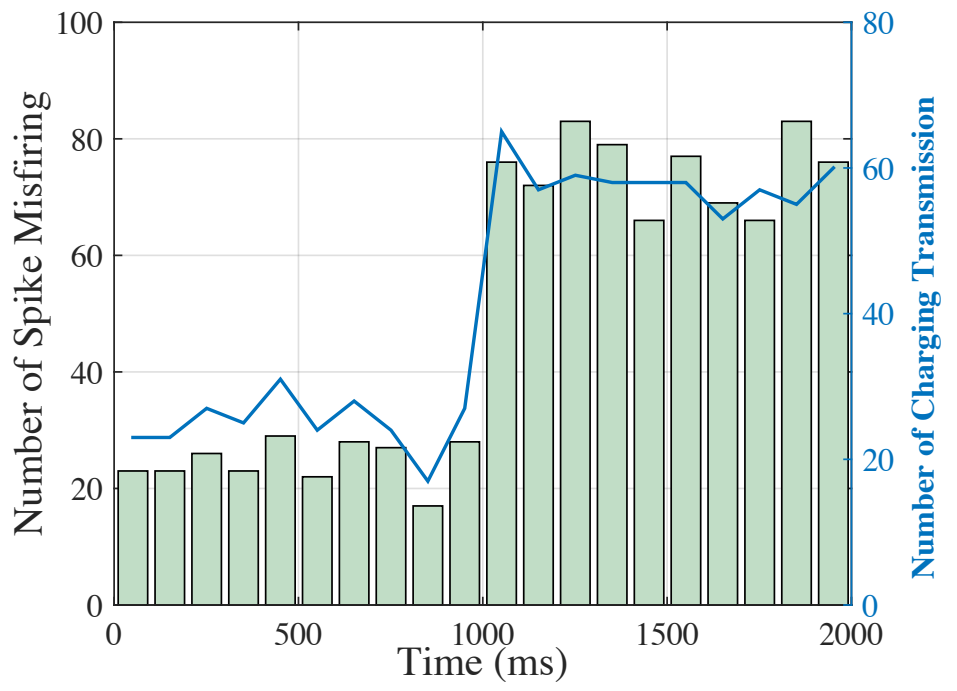

(c)

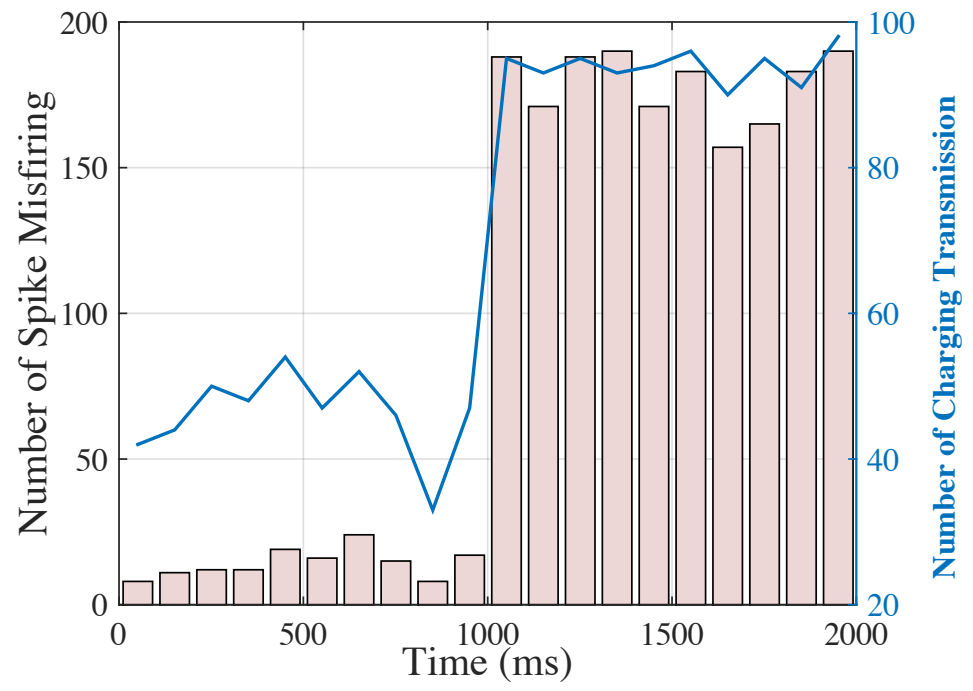

(d)

Fig. 16: (a) The illustration of bi-directional stimuli separated by $60^o$ for achromatic random-dots pattern that is visually observed by a *macaque monkey*. (b) The raster plot simulation generated based on the experiment. As shown in the raster plot, both directions affect the neuron spike frequency response [29]. (c) and (d) presents the simulation results from the number of misfiring before and after the frequency transition for both the *Predictive Sliding Detection Window* and the *Charge and Fire* protocols.

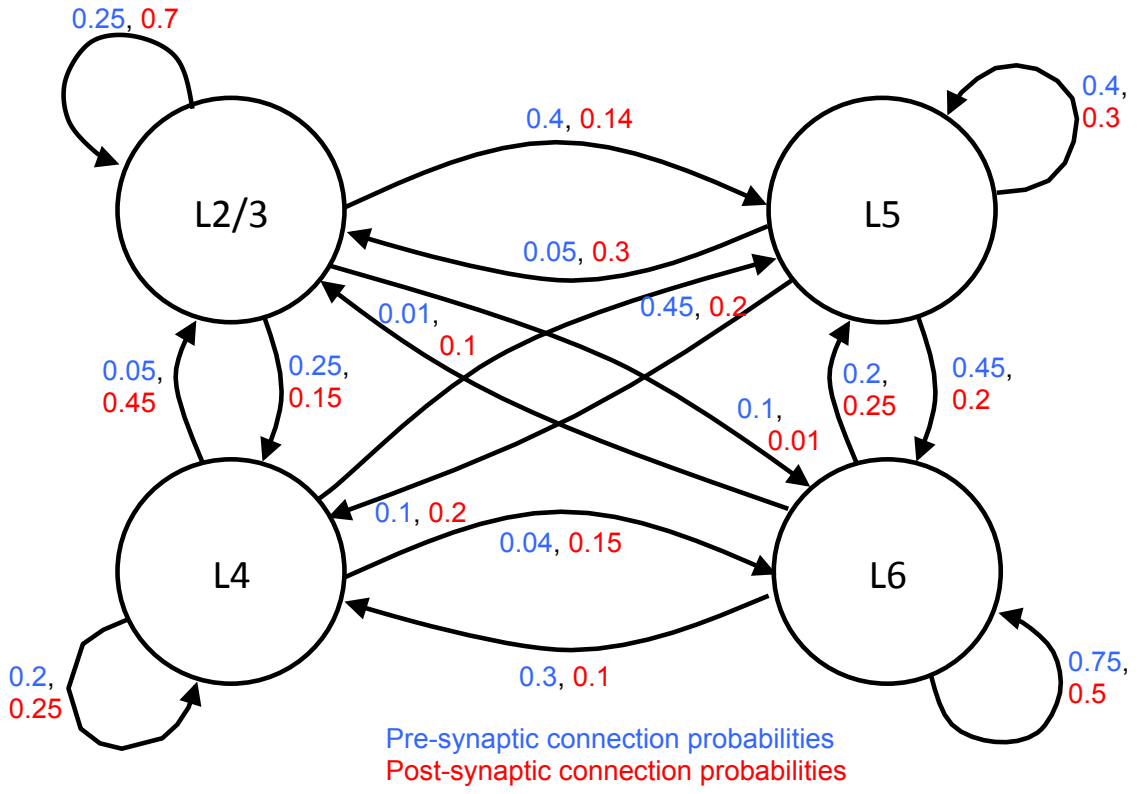

Fig. 17: Markov Chain model of inter and intra-layer connectivity for the cortical column.

TABLE IV: Rank and Connection pattern.

| Rank | Connection Patterns |
|---|---|
| 1 | $L[5] \rightarrow L[6] \rightarrow L[4] \rightarrow L[2/3]$ |
| 2 | $L[4] \rightarrow L[5] \rightarrow L[6] \rightarrow L[2/3]$ |
| 3 | $L[2/3] \rightarrow L[6] \rightarrow L[5] \rightarrow L[4]$ |
| 4 | $L[6] \rightarrow L[4] \rightarrow L[5] \rightarrow L[2/3]$ |
| ⋮ | |
| 24 | $L[6] \rightarrow L[2/3] \rightarrow L[4] \rightarrow L[5]$ |

excitation that occurs in layers with high connection distribution, which will result in the WiOptNDs of that layer to be discharged more frequently. Therefore, the selection of these layers with high centrality for charging will highly depend on the probability of connectivity between the layers as illustrated in Fig. 17, and this particular property can be utilized to improve the *Predictive Sliding Detection Window* protocol.

We can calculate the distribution of a connection either entering to a layer (pre-synaptic synapse - $Pr_{pre}$) or leaving a layer (post-synaptic synapse - $Pr_{post}$). For the Markov chain depicted in Fig. 17, the difference between the two is in the transition probabilities. The probability of the connection distribution $L[y]$ for layer $y$ can be represented as:

$$Pr(L[y]) = E[Pr_{pre}(L[y]) + Pr_{post}(L[y])], \quad (19)$$

where $E[.]$ is the expected value, $P_{pre}(.)$ is the probability of a pre-synaptic connection and $P_{post}(.)$ is the probability of a post-synaptic connection for a layer. These probabilities can be calculated in the same way, but ultimately the behaviour is different due to the transition probability values. Therefore,

$$Pr_{(.)}(L[y]) = \frac{\sum_{k=0}^{|L|} Pr(L[k]|L[y])}{\sum_{n=0}^{|L|}\sum_{m=0}^{|L|} Pr(L[n]|L[m])}. \quad (20)$$

Using the $Pr(L[y])$, the predefined time-delay patterns can be adjusted based on the max$\{\forall Pr(L[y])\}$. According to the number of cortical layers, there are $|L|!$ possible patterns to be selected from. The connection probability is summarized in Table III. An example of a single connection comparison is between $Pr(L[5] \rightarrow L[6]) = 0.325$ and $Pr(L[5] \rightarrow L[4]) = 0.15$, which means that the connection flow from layer V to layer VI is more probable than to layer IV. For this reason, in

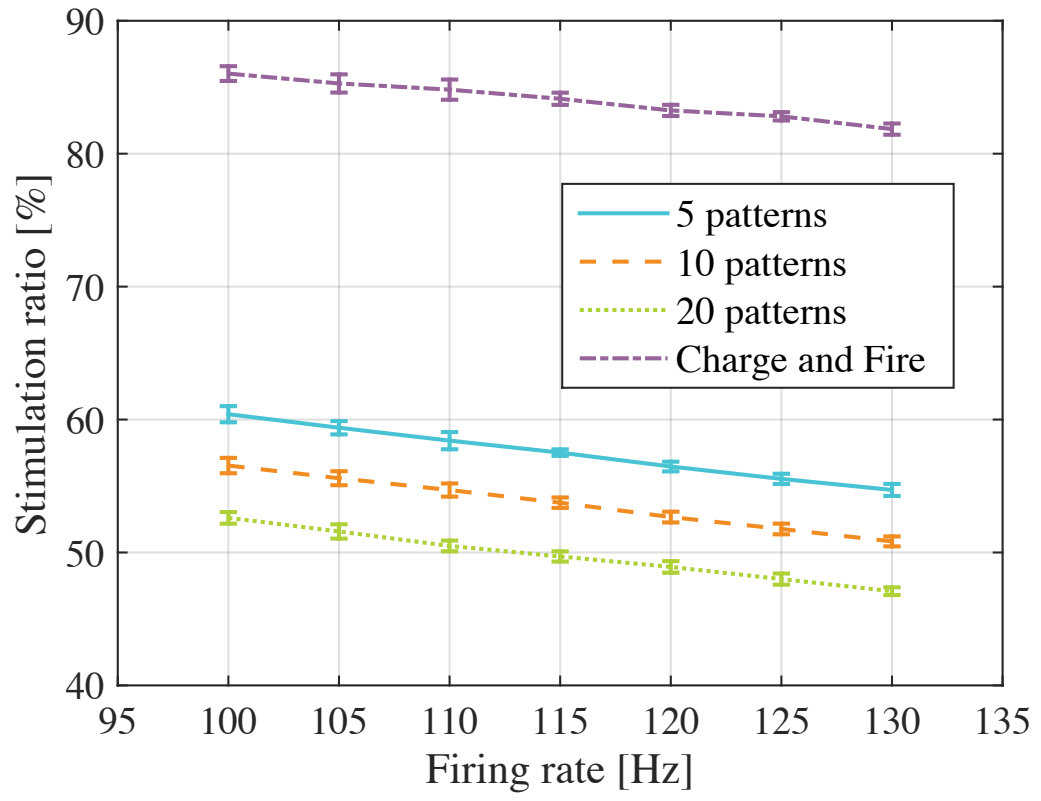

(a) Stimulation ratio vs Spike frequency.

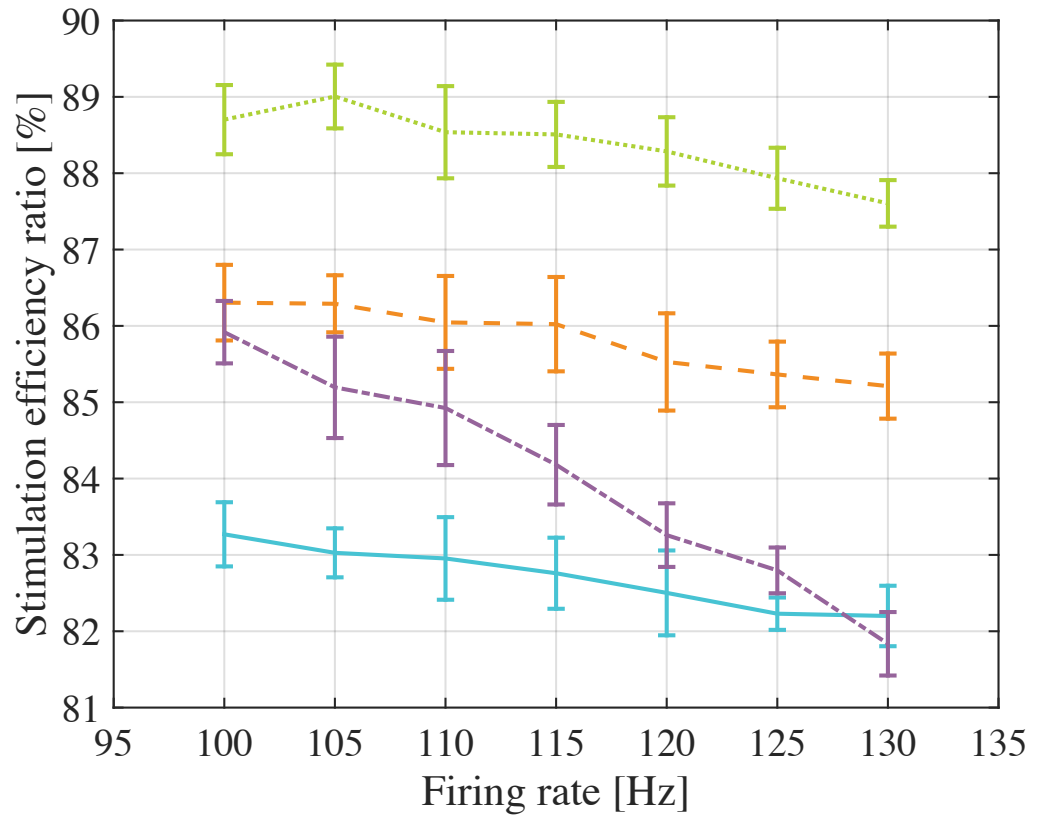

(b) Stimulation efficiency ratio vs Spike frequency.

Fig. 18: Comparison of the stimulation and efficiency ratio of predefined patterns for the *Markov-Chain Time Delay Patterns* and the *Charge and Fire* protocols. The simulation generates the average and standard deviation values of devices deployed in four layers of the cortical column of the cerebral cortex.

the connection ranking process, a connection from Layer V to layer VI will be placed higher. Table IV presents a partial table of the ranks for all the feasible connection patterns of the four layers in the cortical column. The combination of connection chain probability is used for the ranking process. Based on the analysis, the pattern of $L[5] \rightarrow L[4] \rightarrow L[6] \rightarrow L[2/3]$ is ranked the highest, while the $L[6] \rightarrow L[2/3] \rightarrow L[4] \rightarrow L[5]$ is ranked the lowest. By listing all the $|L|! = 4! = 24$ possible connections (Table IV), each predicted sequence patterns can be assigned to the available frequencies. For a frequency set of $F = \{f_1, f_2, f_3, ..., f_i\}$, there are $i$ individual frequencies which can be mapped to $i$ ranked sequences in the list.

Based on the predefined connection ranking table that is used to define the pattern predictions for selected frequencies, simulations in MATLAB was conducted and the results are presented in Fig. 18. The stimulation ratio shows stable decreasing trend with respect to the neuronal spiking rate. However, lower stimulation ratio does not always translate to higher efficiency if it results in higher number of spike misfiring. This is reflected in the stimulation efficiency ratio result.

The efficiency ratio of *Charge and Fire* protocol experiences steep decrement as the the firing rate increases, where we can see that when the firing rate reaches 130 $Hz$, the *Markov-Chain based Time Delay Patterns* with 5 patterns starts to outperform the *Charge and Fire* protocol. This is due to the nature of the sliding detection window protocol in targeting several firing sequences of the neurons based on low number of ultrasound frequency charging. Compared to the randomly chosen pattern simulation (Fig. 14), the stimulation efficiency is higher with smaller standard deviation showing consistent improvement.

## VII. Conclusions

The increased attention towards brain stimulation has attracted researchers to search for innovative solutions that can enable long-term deployment as well as the design of miniaturized devices that can self-generate power. At the same time, the emergence of optogenetics has provided a new approach for precise stimulation at the single neuron level. In this paper, we propose the WiOptND that is constructed from nanoscale components and can be embedded into the cortex of the brain to stimulate neurons using light. A thorough description of the circuitry, as well as the components, are presented, including mechanisms of generating power through ultrasonic wave vibrations. The paper presented simulation results on the behaviour of optical light transmission and its effect on the brain tissue, as well as the energy performance of the device based on variations of ultrasonic frequencies and circuitry devices (e.g. capacitors and piezoelectric nanowire area). A number of charging protocols have also been evaluated ranging from the simple *Charge and Fire* to the *Predictive Sliding Detection Window*, and its variant the *Markov-Chain Time Delay Patterns*. The *Predictive Sliding Detection Window* utilizes predicted patterns of ultrasound frequencies to match to the neural spike patterns. The difference between the protocols is to improve energy efficiency by lowering the number of ultrasound emissions from the sub-dura transceiver while ensuring that neural spike misfiring ratio is low. In particular, the *Markov-Chain Time Delay Patterns* extension protocol resulted in the best performance. However, when the efficiency ratio is considered, the protocol highly depends on the neural spike rate and number of applied predicted patterns. The results from our simulation study have demonstrated the feasibility of using the WiOptND nanonetworks for long term deployments in the brain in order to stimulate neurons and provide new approaches for treating neurological diseases.

## Acknowledgment

This work is supported by the Academy of Finland Research Fellow program under project no. 284531. This work has also been supported by the European Union Horizon 2020 CIRCLE project under the grant agreement No. 665564. This publication has also emanated from research supported in part by the Science Foundation Ireland (SFI) CONNECT research centre, which is co-funded under the European Regional Development Fund under Grant Number 13/RC/2077. This work was also supported by the U.S. National Science Foundation (NSF) under Grants No. CBET-1555720.